\documentclass[letterpaper,twocolumn,10pt]{article}
\usepackage{usenix2019_v3}

\usepackage{tikz}
\usepackage{pgfplots}
\usepackage[inkscapearea=page]{svg}
\svgpath{{img/}}
\usepackage{amsmath}
\usepackage{algorithm}
\usepackage{titlesec}
\usepackage[noend]{algpseudocode}
\usepackage{pdfpages}
\usepackage{filecontents}

\usepackage[font=small,labelfont=bf]{caption}

\titleformat{\subsection}
  {\normalfont\normalsize\bfseries}{\thesubsection}{1em}{}
\titlespacing*{\subsection}{0pt}{1em}{1em}

\titleformat{\subsubsection}[runin]
  {\normalfont\normalsize\bfseries}{\thesubsubsection}{1em}{}
\titlespacing*{\subsubsection}{0pt}{0.5em}{1em}

\begin{document}

\date{}

\title{\Large \bf{Fast ACS: Low-Latency File-Based Ordered Message Delivery at Scale}}
\author{Sushant Kumar Gupta\qquad Anil Raghunath Iyer\qquad Chang Yu\qquad Neel Bagora\\Olivier Pomerleau\qquad Vivek Kumar\qquad Prunthaban Kanthakumar\\\\Google LLC\\\\fast-acs-wp@google.com}

\maketitle

\begin{abstract}
Low-latency message delivery is crucial for real-time systems. Data originating from a producer must be delivered to consumers, potentially distributed in clusters across metropolitan and continental boundaries. With the growing scale of computing, there can be several thousand consumers of the data. Such systems require a robust messaging system capable of transmitting messages containing data across clusters and efficiently delivering them to consumers. The system must offer guarantees like ordering and at-least-once delivery while avoiding overload on consumers, allowing them to consume messages at their own pace.

This paper presents the design of Fast ACS (an abbreviation for Ads Copy Service), a file-based ordered message delivery system that leverages a combination of two-sided (inter-cluster) and one-sided (intra-cluster) communication primitives—namely, Remote Procedure Call and Remote Memory Access, respectively—to deliver messages. The system has been successfully deployed to dozens of production clusters and scales to accommodate several thousand consumers within each cluster, which amounts to Tbps-scale intra-cluster consumer traffic at peak. Notably, Fast ACS delivers messages to consumers across the globe within a few seconds or even sub-seconds (p99) based on the message volume and consumer scale, at a low resource cost.
\end{abstract}

\section{Introduction}

Low-latency message delivery is crucial for real-time systems such as advertising, dynamic pricing, retail inventory management, fraud detection, activity monitoring, online matching, and online gaming. The explosion of data is driving a corresponding growth in the size and complexity of real-time systems. These systems are often geographically distributed across numerous clusters. This distribution is essential for handling the immense volume of requests and maintaining a low-latency response. Furthermore, these systems are designed to scale horizontally, which is vital because they often handle internet-scale traffic with user numbers that can fluctuate dramatically. This also makes them vulnerable to internet-scale threats like DDoS attacks.

Consider the example of Google Ads, a large-scale online advertising system. The Ads serving system is geographically distributed across multiple clusters to handle massive user traffic. Each cluster runs numerous jobs with multiple replicas to ensure high availability and low response time. Some jobs may have thousands of replicas per cluster, while others may have tens of thousands, resulting in hundreds of thousands of replicas globally. Another example of a large-scale real-time system is online shopping serving. They need to be highly scalable to accommodate peak shopping periods when demand can increase dramatically. This requirement for dynamic scalability is common to many real-time systems that serve internet-scale traffic.

Such real-time systems rely on real-time data pipelines to receive updates for their states. These pipelines utilize real-time extractors~\cite{zaharia2013discretized,akidau2013millwheel,carbone2015apache,akidau2015dataflow} that depend on a robust message delivery subsystem. This subsystem transports messages containing extracted data to the real-time systems. Depending on the specific needs of the real-time system, the message delivery subsystem must offer certain guarantees. For example, stateful real-time systems that can handle duplicate messages require \emph{in-order} sequencing (messages delivered in the order they were produced) and \emph{at-least-once} delivery (each message delivered at least once to all consumers). Ordered delivery encompasses unordered delivery and at-least-once delivery subsumes at-most-once delivery. While exactly-once delivery is desirable too, it requires two-phase commit, which is challenging for cross-cluster delivery and hence beyond the scope of this paper. Beyond delivery guarantees, low latency is paramount for these messaging systems, as is the ability to handle a massive number of consumers.

\subsubsection*{Existing Message Delivery System}

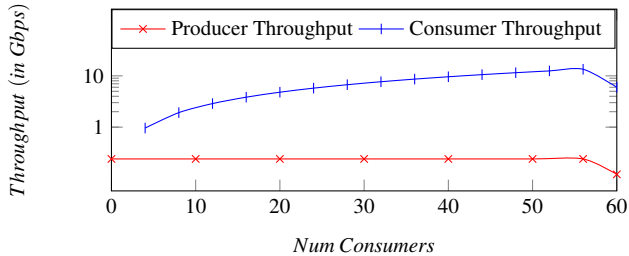
\begin{figure}
\fontsize{8}{10}\selectfont
\begin{tikzpicture}
\begin{axis}[
    xlabel=$Num\ Consumers$,
    ylabel=$Throughput\ (in\ Gbps)$,
    xmin=0, xmax=60,
    ymin=0, ymax=200,
    xtick={0,10,...,60},
    ytick={0,1,10},
    yticklabels={1,10},
    ymode=log,
    height=4cm,
    width=8.3cm,
]

\addplot[smooth,red,mark=x] plot coordinates {
    (0,0.24)
    (10,0.24)
    (20,0.24)
    (30,0.24)
    (40,0.24)
    (50,0.24)
    (56,0.24)
    (60,0.12)
};

\addplot[smooth,blue,mark=|] plot coordinates {
    (0,0)
    (4,0.96)
    (8,1.92)
    (12,2.88)
    (16,3.84)
    (20,4.8)
    (24,5.76)
    (28,6.72)
    (32,7.68)
    (36,8.64)
    (40,9.6)
    (44,10.56)
    (48,11.52)
    (52,12.48)
    (56,13.44)
    (60,5.984)
};

\pgfplotsset{legend columns=-1,legend style={at={(0,1)},anchor=north west}}
\addlegendentry{Producer Throughput}
\addlegendentry{Consumer Throughput}

\end{axis}
\end{tikzpicture}
\caption{Results for Kafka consumer-scaling experiment. As consumers scaled, the consumer throughput went up to 14 Gbps, after which the performance degraded.}
\vspace{-1.5em}
\label{fig:expkafka}
\end{figure}

Message delivery systems are available in two variants: \emph{push-based} and \emph{pull-based}. Pioneered by systems like Scribe~\cite{rowstron2001scribe}, Siena~\cite{carzaniga2000achieving}, and Gryphon~\cite{strom1998gryphon}, push-based asynchronous message delivery has found widespread adoption in RabbitMQ~\cite{dossot2014rabbitmq}, Facebook's Wormhole~\cite{sharma2015wormhole}, Google Cloud PubSub~\cite{krishnan2015google}, and Apache ActiveMQ~\cite{christudas2019activemq}. However, relying on a push-based model presents challenges for many systems. When a diverse set of consumers, with their varying latency-sensitive backends, needs to be supported, push-based delivery becomes impractical. Handling push messages can block critical CPU cores, leading many systems to prefer pulling messages at their own pace. Moreover, pull-based messaging systems can deliver higher throughput that push-based systems.


Systems such as Apache Kafka~\cite{kreps2011kafka} have implemented pull-based models, employing brokers to store messages in files on their local file system. Each broker handles one or more \emph{partitions} of a \emph{topic}. Within a partition, messages are totally ordered, while across partitions, they are partially ordered. The throughput of brokers, for both writing and reading, is constrained by the capabilities of the storage medium (like SSDs, currently maxing out at about 4.84 Gbps). Kafka performs well when message volume is high (in the Gbps range) and consumer count is limited. However, it is not optimized for scenarios with a large fan-out; tens of thousands of consumers can lead to potential meltdown, at the very least cause a significant increase in latency. There are ongoing efforts to replace local file system with Hadoop Distributed File System (HDFS)~\cite{shvachko2010hadoop}. However, low-latency delivery to several consumers requires enormous tail-reading of file bytes, and so HDFS data node servers with a 64 MB file block size can still face throughput issues. In-memory caching of "hot" messages within brokers offers a potential workaround for storage bottlenecks. However, hot-spotting and network congestion can occur even within a single partition, for instance, when global total order is required.

To demonstrate consumer throughput issues, we conducted an experiment on Google Cloud~\cite{GoogleCloud}. The experiment involved a Kafka cluster with the following specifications: 3 brokers in a zone, each configured with 8 vCPUs, 32 GB of RAM, and NVMe SSD; and one topic with 9 non-replicated partitions.The partitioning ensured even distribution of load across brokers, preventing hot-spotting. A producer was initiated, sending 1 KB-sized messages at a rate of 240 Mbps. Then, consumers were scaled up incrementally, adding four at each step. Figure~\ref{fig:expkafka} shows the results. Initially, consumer throughput scaled linearly, reaching a peak of 14 Gbps. However, beyond this point, we observed a drop in the write rate due to throughput conflict between writes and reads. The producer byte rate decreased to 120 Mbps, consequently causing a drop in overall consumer throughput. The maximum consumer throughput achieved in this Kafka setup was 14 Gbps, after which performance degraded.

Apache Pulsar~\cite{kjerrumgaard2021apache} is a geo-replicated message delivery system. It uses Apache BookKeeper~\cite{BookKeeper} for storage. BookKeeper persists messages in distributed write-ahead logs, and indexes them in a cache for fast retrieval. These logs are segmented to optimize performance. However, their large segment size (typically 128 MB) can lead to throughput issues when several consumers tail-read from them.

\subsubsection*{Our Contributions}

To the best of our knowledge, there are no messaging systems that can handle large consumer fan-out and provide low-latency ordered message delivery. Our contribution lies in the development of such a system.

In summary, our system utilizes files on a distributed file system, namely Colossus~\cite{Colossus} (formerly Google File System~\cite{ghemawat2003google}). Producers append messages to these files, while consumers poll and read them. We introduce an in-memory layer on top of Colossus specifically designed for low-latency message delivery at scale. The system leverages one-sided network primitives, namely Remote Memory Access (RMA), for intra-cluster reads, while utilizing traditional two-sided communication, namely Remote Procedure Call (RPC), for inter-cluster byte transfer. The system provides essential guarantees for message delivery: \emph{in-order} sequencing and \emph{at-least-once} delivery to all consumers. Additionally, it enables consumers to pull messages at their own pace.

Our contributions are fourfold: 1) We demonstrate how RMA-based in-memory caching overcomes tail-read throughput limitations, enabling low-latency message delivery. 2) We show that proper chunking and distribution eliminate hot-spotting, even with global total ordering constraints. 3) We demonstrate that scaling out resources horizontally effectively mitigates network limitations. 4) We highlight the resource efficiencies achieved by leveraging RMA.

Our contribution targets real-time systems with high consumer throughput requirements, rather than systems that need extremely low latency, say, under 50ms. While achieving such low latency is possible with enhanced resources, it can be expensive. The ideas discussed in this paper can be adapted to enhance the consumer scalability of existing systems, leading to improvements in their benchmark performance.

We have materialized the design to build Fast ACS (an abbreviation for Ads Copy Service). Fast ACS has been deployed in production, consistently delivering messages to thousands of consumers across dozens of production clusters within a few seconds or even sub-seconds (p99). Fast ACS is adopted by systems that demand ordering, at-least-once delivery and low-latency on a large scale.

\begin{figure*}[htbp]
\fontsize{8}{10}\selectfont
\centering
\def\svgwidth{17cm}
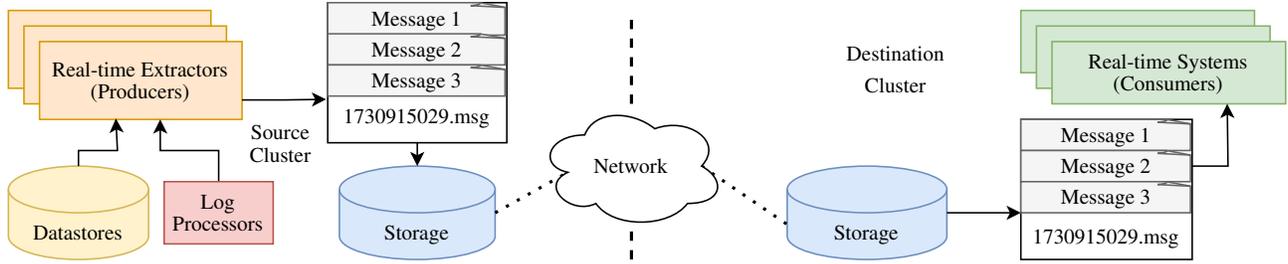
\caption{\label{fig:introduction} File-based ordered messaging for a real-time systems. Each message stream can have multiple destination clusters based on its consumers. Message files are tail-copied from the source cluster's storage to the destination clusters' storage. }
\vspace{-1.5em}
\end{figure*}

\section{Design Overview}
\label{sec:design}

We will begin by outlining how our system achieves ordered message delivery through files. Starting from Section~\ref{sec:storage}, we will detail our system's core components.

\subsection{File-Based Ordered Message Delivery}

Files provide a mechanism for ordered message delivery. Producers append messages to \emph{message files} and consumers read from them. To limit their size, the message files undergo periodic rollover. Each message file includes its creation timestamp in its name, enabling chronological sorting.

This file-based messaging approach is agnostic to the underlying message formats, focusing primarily on the storage and delivery of bytes. Consequently, producers and consumers can define their own protocols for serializing and deserializing messages to and from the files.

Producers write messages into distinct \emph{message streams} in the source cluster, each representing an individual bundle of messages, analogous to \emph{topics} in pub-sub systems. Producers can write to multiple message streams, while consumers can read from multiple ones as well. Each message stream is sharded for scalability, with each shard comprising a set of chronological message files containing the messages. A message stream shard resides within a single directory of the file system, simplifying organization. Thus, all ordered messages for a single message stream shard can be found in chronological files under one directory.

To accommodate geographically distributed consumers, message files must be copied from the source cluster's storage to the destination cluster's storage. Consumers in the destination cluster can then tail-read the file from their cluster-local storage system. Figure~\ref{fig:introduction} provides an illustration of our cross-cluster file-based ordered message delivery for a real-time system.

\begin{figure}[htb]
\fontsize{8}{10}\selectfont
\centering
\def\svgwidth{\columnwidth}
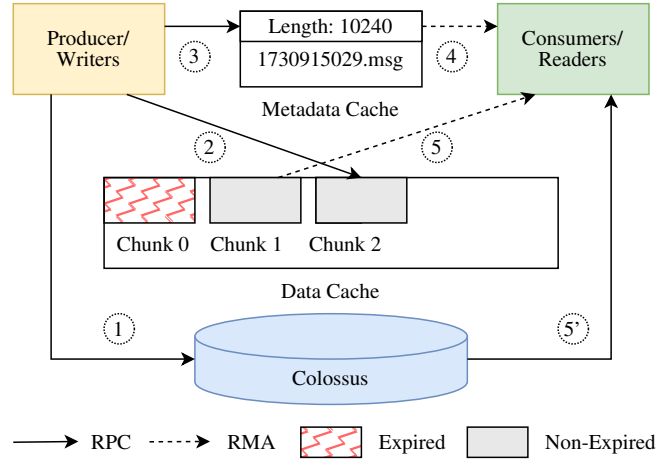
\caption{\label{fig:design} The multi-layer storage design. The producers/writers write the bytes to the Colossus file (1) and shadow the bytes as chunks to the data cache (2), and then update the length of the file in the metadata cache (3). The consumers/readers poll the metadata cache for the latest length of the file (4) and read the chunks from the data cache preferably (5), with Colossus acting as a fallback (5'). All the reads from caches happen over RMA. In this  illustration, chunk 0 has expired while chunk 2 is only partially-filled. The length stored in the metadata cache points to the current position within this last, partially-filled chunk.}
\vspace{-1.5em}
\end{figure}

\subsection{Multi-Layer Storage}
\label{sec:storage}

As illustrated in Figure~\ref{fig:design}, the message files are stored in a multi-layered storage system: Colossus~\cite{Colossus}, a distributed file system, serves as the primary layer, while a replicated in-memory cache acts as the secondary layer. This additional layer ports minimal Colossus features and operates in parallel to it. All message stream bytes are replicated to both layers. Given its finite capacity, the in-memory cache layer only stores the most recently written, "hot" bytes for the files. The layer absorbs the majority of queries for tail-reads, with Colossus serving as a fallback. This offers a low-latency, high-throughput path for message delivery.

\subsubsection*{The In-Memory Cache Layer} Files are segmented into a series of fixed-size chunks, each 4 KB in size. These chunks are then represented as key-value pairs. The key for each chunk is a hash of two components - the absolute Colossus file path, which is globally unique across Colossus, and the chunk's sequence number. The value comprises the corresponding chunk bytes. The 4 KB chunk size is partly based on the observed message size and partly on the maximum transmission unit in the cluster fabric, allowing a chunk to be read in a single packet.

We employed CliqueMap~\cite{singhvi2021cliquemap} as the underlying system for the in-memory storage layer. CliqueMap is a remote in-memory key-value cache that facilitates writes through RPC and intra-cluster reads via RMA. The latter proves advantageous given the read-heavy nature of our intra-cluster consumer traffic. Each key-value pair in the cache is subject to garbage collection (GC) upon either the expiration of its Time-To-Live (TTL) or when the cache approaches its storage capacity limit. CliqueMap employs consistent hashing~\cite{karger1997consistent} for key distribution among replicas. Within a replica, CliqueMap divides its memory into fixed-size \emph{slabs}, and key hashes are divided into \emph{buckets} via set-associative mapping.

CliqueMap supports r=3.2 replication, which provides resilience against single-point failures. With replication, CliqueMap offers a range of read consistency modes, two of which we leverage: the \emph{consistent} mode, which reads values from at least 2 out of 3 replicas, thereby delivering the most recent value, and the \emph{relaxed} mode, which reads from a random replica that may provide a stale value.

In constructing a file system atop CliqueMap, we employ two instances per cluster: a \emph{data cache} and a \emph{metadata cache}. These are analogous to the Colossus \emph{chunkservers} and \emph{metadata servers} respectively, except lacking processing capabilities. As a result, unlike Colossus, the data and metadata for a file cannot be atomically written to the cache layer. The data cache acts as the repository for file chunks while the metadata cache stores file metadata, particularly (a) the current \emph{length} of the file and (b) the \emph{file lock}. The current length, a monotonically increasing number, reflects the file length of the append-only file, stored as chunks, in the data cache, while the file lock serves to prevent concurrent writers in a best-effort manner. As elaborated subsequently, locks primarily contribute to performance optimization and do not impact the system's correctness. While both file chunks and metadata stored in such volatile caches remains vulnerable to data loss, our system is designed to withstand such occurrences, as detailed in the subsequent sections.

The data cache has two types of chunks for each file: \emph{complete} chunks and a \emph{last, partially filled} chunk (if any). Each chunk in the data cache possesses the following property: reading any part of a chunk will always yield the correct bytes, if present, for the corresponding file positions, even if the chunk is partially filled. This is guaranteed by always writing a chunk starting from its beginning and never from any other position, as detailed later. TTL expiration automatically removes older, non-hot chunks.

\begin{figure}[htbp]
\fontsize{8}{10}\selectfont
\centering
\def\svgwidth{\columnwidth}
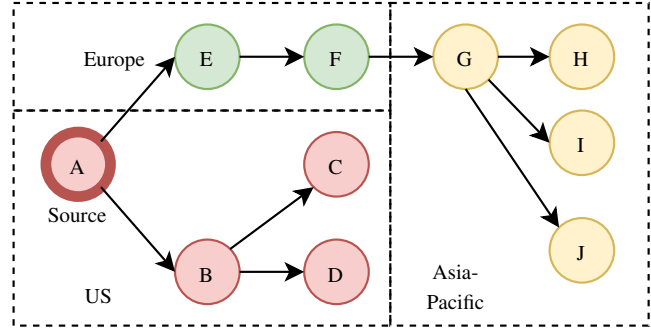
\caption{\label{fig:copy_tree} An MST for a message stream. The source cluster is in the US, and the stream is copied to three clusters within the US, two clusters in Europe, and four clusters in the Asia-Pacific region. }
\vspace{-1.5em}
\end{figure}

\subsection{Routing}

Directly streaming bytes from the source cluster to all the destination clusters of a message stream would be prohibitively expensive. To address this, we utilize a Prim-based~\cite{prim1957shortest} optimizer to construct a minimum spanning tree (MST),
hereafter referred to as a \emph{copy tree}, for each message stream, as illustrated in Figure~\ref{fig:copy_tree}. While prioritizing bandwidth cost, the optimizer is also fine-tuned to minimize both, the tree's depth and the fan-out per node. Minimizing the tree's depth is essential, as each hop along the copy tree introduces some delay. The first hop from source typically traverses continental boundaries, followed by a hop across metro boundaries. A typical copy tree has a maximum of 4 hops from the source cluster which can accommodate a few hundred clusters.

Outages in a cluster, often caused by factors such as bad releases, connectivity problems, or bulk maintenance events, can delay all descendant clusters in the copy tree. Such disruptions frequently led to significant Service-Level Objective (SLO) misses, which prompted us to automate the outage handling process. We employ a breadth-first search algorithm to penalize the problematic nodes and links, and then delegate reconstruction of the copy tree to the optimizer. The reconstruction usually sets the problematic nodes as leaf nodes in the copy tree. Consumers in the affected cluster are re-routed to a nearby healthy cluster, sacrificing the ability to perform RMA reads. Similarly, when a new cluster is added to a message stream, it starts as a leaf node with rate-limited delivery and is then promoted to a regular node once the catch-up to the head is complete.

Along each hop of the copy tree, we have a pair of worker jobs - \emph{readers} and \emph{writers}. Readers are responsible for polling, reading, and sending bytes from the current hop to the next hops of the copy tree while writers write the bytes to the destination storage. This design provides the necessary decoupling for the jobs to scale independently. The readers scale based on the egress bandwidth demands, which depends on the fan-out from the current cluster, and the writers scale based on the ingress bandwidth demands.

Each file is tail-copied independently through an \emph{operation}, identified by the Colossus target file path and the storage type - either Colossus or cache. As a result, for each file, the operations for Colossus and cache are independent along the entire copy tree, which also doubles the bandwidth utilization. While pairing them was technically feasible, it led to additional complexity in implementation and introduced avoidable latency. For example, invoking the Colossus \emph{Open} file function in the write mode is prohibitively expensive as it needs to acquire lock on the file. The cache operation bypasses such costly Colossus operations and is heavily-optimized for parallelization, enabling the fastest possible byte delivery rate.

Finally, operations are scheduled by a cluster-local \emph{scheduler}, which is triggered by a notification from the producer. The schedulers establish a continuous stream along the tree.

\subsection{Components}

\subsubsection*{Producers} Producers (i.e., real-time extractors) are responsible for generating update files and appending serialized messages to them. Our \emph{File} interface implementation writes the serialized messages to the underlying Colossus file and then asynchronously shadows (copies) the bytes to the data cache. The length of the Colossus file and the length of the file in the data cache are then recorded in the metadata cache. This allows for low-latency retrieval of the file lengths from the metadata cache via RMA, especially for Colossus, where Colossus Bigtable~\cite{chang2008bigtable} often runs into hot-spotting issues when several consumers poll it simultaneously.

Flushing messages to Colossus establishes a single source of truth, with all other copies (in other Colossus clusters and caches) reflecting this committed state. To amortize disk write costs, producers typically buffer messages before flushing.

In order to shadow the bytes to the data cache, the file implementation tracks the last partially-filled chunk (if any) for the append-only file. When new bytes are flushed to Colossus, the file implementation computes the chunks for the data cache and performs a bulk-write operation on the corresponding key-value pairs. Since a write to a chunk is atomic, slow-growing files often end up overwriting the same bytes to the same chunk multiple times, thereby increasing intra-cluster bandwidth usage unnecessarily. This can be partially resolved by reducing the chunk size to fit the file's byte rate.

\begin{figure}[htb]
\fontsize{8}{10}\selectfont
\centering
\def\svgwidth{8cm}
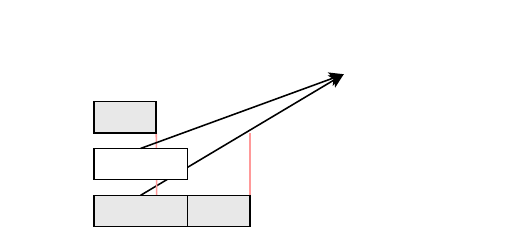
\caption{\label{fig:consistent_read} An illustration of relaxed reads followed by consistent reads. Suppose new bytes are written across multiple chunks, filling up the previously partially-filled chunk 0. The new chunk 1 is successfully written to all replicas, but the updated chunk 0 is only written to replicas 2 and 3. When a reader detects a new length from the metadata cache and performs a relaxed read for chunk 0 from replica 1, it will receive fewer bytes than expected. The reader will then perform a consistent read to obtain the complete chunk 0.}
\fontsize{8}{10}\selectfont
\centering
\def\svgwidth{7cm}
\begingroup%
  \makeatletter%
  \providecommand\color[2][]{%
    \errmessage{(Inkscape) Color is used for the text in Inkscape, but the package 'color.sty' is not loaded}%
    \renewcommand\color[2][]{}%
  }%
  \providecommand\transparent[1]{%
    \errmessage{(Inkscape) Transparency is used (non-zero) for the text in Inkscape, but the package 'transparent.sty' is not loaded}%
    \renewcommand\transparent[1]{}%
  }%
  \providecommand\rotatebox[2]{#2}%
  \newcommand*\fsize{\dimexpr\f@size pt\relax}%
  \newcommand*\lineheight[1]{\fontsize{\fsize}{#1\fsize}\selectfont}%
  \ifx\svgwidth\undefined%
    \setlength{\unitlength}{203.25bp}%
    \ifx\svgscale\undefined%
      \relax%
    \else%
      \setlength{\unitlength}{\unitlength * \real{\svgscale}}%
    \fi%
  \else%
    \setlength{\unitlength}{\svgwidth}%
  \fi%
  \global\let\svgwidth\undefined%
  \global\let\svgscale\undefined%
  \makeatother%
  \begin{picture}(1,0.44280443)%
    \lineheight{1}%
    \setlength\tabcolsep{0pt}%
    \put(0,0){\includegraphics[width=\unitlength,page=1]{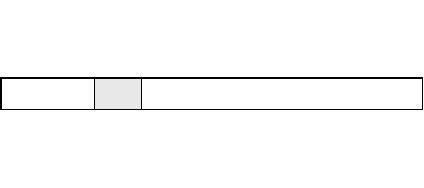}}%
    \put(0.49704796,0.37084871){\makebox(0,0)[t]{\lineheight{1.25}\smash{\begin{tabular}[t]{c}Read Range from Data Cache\end{tabular}}}}%
    \put(0,0){\includegraphics[width=\unitlength,page=2]{fallback_read.pdf}}%
    \put(0.67416973,0.03874539){\makebox(0,0)[t]{\lineheight{1.25}\smash{\begin{tabular}[t]{c}Fallback Read Range from Colossus\end{tabular}}}}%
    \put(0,0){\includegraphics[width=\unitlength,page=3]{fallback_read.pdf}}%
    \put(0.11808118,0.03874539){\makebox(0,0)[lt]{\lineheight{1.25}\smash{\begin{tabular}[t]{l}Missed\end{tabular}}}}%
    \put(0,0){\includegraphics[width=\unitlength,page=4]{fallback_read.pdf}}%
    \put(0.11808118,0.11254613){\makebox(0,0)[lt]{\lineheight{1.25}\smash{\begin{tabular}[t]{l}Read\end{tabular}}}}%
  \end{picture}%
\endgroup%

\caption{\label{fig:fallback_read} An illustration of fallback reads to Colossus in the event of misses from the data cache. All missed chunks are fetched in a single Colossus read to amortize the RPC cost. }
\vspace{-1.5em}
\end{figure}

\subsubsection*{Readers} For each operation, a reader replica in an upstream cluster initiates a long-lived streaming RPC connection~\cite{gRPC} with a writer replica in a downstream cluster. The operation polls the file systems for file length changes and subsequently reads and sends the bytes downstream.

File length is periodically polled from the metadata cache, which stores lengths for both the data cache file and the Colossus file. To reduce overhead, a single background thread performs consistent bulk reads from the metadata cache on behalf of all the operations scheduled on the reader replica. The polling interval for the metadata cache is set to 50ms, and can be reduced to 10ms or less based on latency requirements. Additionally, the Colossus file length is also polled from Colossus Bigtable, albeit less frequently. This serves as a fallback, in case the metadata cache becomes unavailable.

The data cache reads employ a two-step process. Initially, a relaxed read for chunks is performed, followed by consistent reads for those chunks that return fewer bytes than expected — a scenario that may occur when reading from a lagging data cache replica, as illustrated in Figure~\ref{fig:consistent_read}. Empirical observations show that consistent reads are infrequent. Hence, relaxed reads contribute to saving substantial intra-cluster bandwidth and processing resources. An alternative approach could involve using \emph{2xR}, where a quorum on the version number of the key-value pair is first established. However, this necessitates two round-trips, which introduces latency. The relaxed read is hedged after a short delay of 30ms.

Note that because the data cache is volatile, it is possible to encounter a missing chunk even after a consistent read. A missing chunk is retrieved from Colossus by reading from the positions corresponding to the chunk. In scenarios where multiple chunks are missing, the corresponding read from Colossus spans the combined positions of all the missing chunks as illustrated in Figure~\ref{fig:fallback_read}.

For a Colossus operation, the bytes are retrieved opportunistically from the data cache first and then from Colossus upon a miss, as described above. Opportunistic reads from the data cache in Colossus operations are best-effort but save valuable disk time. The Colossus Flash Cache~\cite{yang2022cachesack} is ineffective as it only caches immutable blocks, which is not useful for tail-reading growing files. The read bytes are relayed to the downstream cluster sequentially, in line with Colossus's append-only operation support. 

\begin{figure}[htbp]
\centering
\def\svgwidth{8cm}
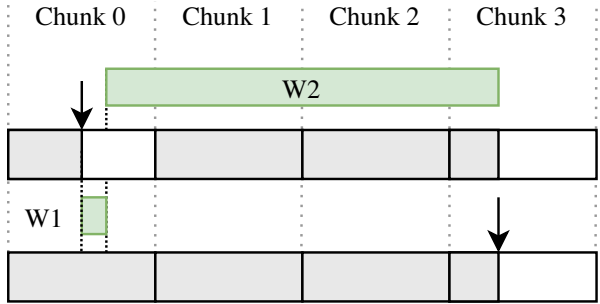
\caption{\label{fig:writes} An illustration of writes to data cache. Consider a scenario where the 0th chunk is partially-filled and there are two concurrent write requests: $W_1$ with a few bytes (not enough to fill the 0th chunk) and $W_2$ with large bytes (spanning over the 0th, 1st and 2nd chunk fully, and the 3rd chunk partially). It is possible for $W_2$ to arrive before $W_1$, leading to the 1st, 2nd chunk and 3rd chunk being written before the 0th chunk. Consequently, the $W_2$th write for the 0th chunk needs to be serialized after $W_1$ is received and the chunk is complete. The length, shown as pointer, is updated only after all bytes up to the pointer are filled.}
\vspace{-1.5em}
\end{figure}

Similarly, for a cache operation, chunks are fetched from the data cache first, followed by a fallback to Colossus upon a miss. However, unlike a Colossus operation, the bytes are read and sent to the writer in parallel, i.e., the processing of the next file length delta does not wait for the previous one to complete.

Astute readers may notice a circular dependency between Colossus and cache operations. However, this circularity is resolved by the fact that actual byte reads are limited to the length of the file in the corresponding storage only. The cache reads by Colossus operations are \emph{opportunity-based} while the Colossus reads by the cache operations are \emph{fallback-based}.

\subsubsection*{Writers}

For a Colossus file, the operation employs single-threaded appends, which also updates the metadata in Colossus Bigtable, constituting an atomic step. 

For the data cache, the operation differentiates between complete chunks and the last partially-filled chunk. Complete chunks are written in parallel as they will only be written once. On the other hand, writes to partially filled chunks are serialized to ensure that the prefix of bytes in the chunk is always correct, as illustrated in Figure~\ref{fig:writes}. This creates per-chunk dependencies between write requests and the operation maintains a map of such write relationships for each chunk. Once all writes are complete, the length in the metadata cache is then updated to reflect the position up to which all complete chunks and the last partially-filled chunk have been written to the data cache. As noted earlier, unlike a Colossus file, the writes to the data and the metadata cache do not constitute an atomic step.

\begin{algorithm}
\caption{Delayed Colossus Reads for a File}\label{dfr}
\begin{algorithmic}[1]
\State $\textit{maxDelay} \gets \textit{1s}$ \Comment{Maximum tolerable read delay.}
\State $\textit{cfsSizeRec} \gets \textit{queue()}$ \Comment{Colossus file size historian.}
\\
\Procedure{ShouldReadFromColossus}{}
\State $\textit{cacheSize} \gets \textit{cache.size()}$
\State $\textit{cfsSize} \gets \textit{colossus.size()}$
\State \textit{cfsSizeRec.push(\{time:\ time(),\ size:\ cfsSize\})}
\\
\While{\textit{!cfsSizeRec.empty()}}
\If {\textit{cfsSizeRec.front().size} > \textit{cacheSize}}
\State \textbf{break}
\EndIf
\State \textit{cfsSizeRec.pop()} \Comment{Remove caught-up positions.}
\EndWhile
\If {\textit{cfsSizeRec.empty()}}
\State \Return false \Comment{Cache is caught-up.}
\EndIf
\If {\textit{time()} > \textit{cfsSizeRec.front().time} + \textit{maxDelay}}
\State \textit{cfsSizeRec.clear()}
\State \Return true \Comment{Cache is delayed.}
\EndIf
\State \Return false \Comment{Cache may catch-up soon.}
\EndProcedure
\end{algorithmic}
\end{algorithm}

Note that this out-of-order parallel-write scheme for the data cache interferes with evictions based on TTL where chunks with higher sequence number may be evicted before chunks with lower sequence number. In practice, this has not been a significant problem because consumers rarely fall behind enough to be impacted by chunk expiration.

\subsubsection*{Consumers} Consumers (i.e., real-time systems) are responsible for reading bytes from the storage and deserializing the messages in a sequential manner. Our design guarantees eventual progress for the consumers, regardless of the volatile state of either the metadata or data caches. This is due to the persistent, append-only nature of Colossus files.

Like readers, consumers actively poll the metadata cache to get the file length for both the Colossus file and the data cache file. Consumers then have the option to read from either of the two storage layers. A naive approach would be to read the chunks from the storage which is ahead of the other in order to minimize latency. Typically, the data cache file leads the Colossus file, making this strategy suitable for most cases. However, we have observed that Colossus files can often overtake the data cache intermittently, particularly when a data cache stream is interrupted along some copy tree hop. This used to result in a significant reads from Colossus, leading to increased latency penalties when Colossus throttled the requests. To address this, our objective was to prioritize reading from the data cache as long as latency requirements are met. This is achieved by postponing reads from Colossus unless message delivery latency is at risk. The implementation for delayed Colossus reads is illustrated in Algorithm~\ref{dfr}. Our \emph{File} interface implementations makes this process transparent to the consumer jobs.

Sometimes, some consumers might start late or fall behind. In such cases, past reads from the data cache will fail as the chunks would have expired, and the consumers are forced to fallback to Colossus until they catch up to the head and can resume reading from the cache. Upon cache miss, the fallback mechanism is same as that described for readers.

Lastly, each consumer passively rate-limits the reads to cap the read bandwidth in the event that the producer restarts and writes a large backlog to the cache. Without the rate limit, the read amplification would overwhelm the data cache.

\section{Implementation Details}
\label{sec:implementation}

\subsubsection*{Start Position for Cache Copies} In essence, the data cache is designed to serve only non-lagging consumers. Consumers who have fallen behind are expected to catch up using Colossus file before reading chunks from the data cache. This simplification allows us to help keep the data cache stream ahead of the Colossus stream. In cases of cache outages and subsequent recovery, the data cache stream jumps to the chunk where the Colossus file is currently positioned, and the length in the metadata cache is updated accordingly.

\subsubsection*{Dueling Writers} The scheduler employs Slicer~\cite{adya2016slicer} for sharding, which, like many sharding schemes, provides weak consistency guarantees. While strong consistency could have been achieved by employing a lock service like Chubby~\cite{burrows2006chubby} or ZooKeeper~\cite{hunt2010zookeeper}, or by utilizing Slicer's strong consistency mode, these approaches compromise availability. Prioritizing availability, we opted for Slicer's weak consistency model. Consequently, multiple operations can exist for the same target path and storage. 

For Colossus files, multiple writers are prohibited since the \emph{Open} operation on a file invalidates all existing file handles used by other processes. This allows a new operation to automatically supersede an older one for a target path. 

However, such a mechanism is not available for cache operations as the cache storage layer cannot execute logic. This can result in multiple writers competing, causing clobbering of key-value pairs in both the data and the metadata caches. For instance, complete chunks would be overwritten with partial chunks or the file length would intermittently go backwards. This would negatively impact performance, forcing consumers to frequently fallback to Colossus. Note that the correctness remains unaffected, and appendix~\ref{sec:appendix} provides a formal specification verifying the system's correctness.

To address the performance issue, we utilize locks stored in the metadata cache. These locks are essentially leases that can be marked as \emph{poisoned}. Each lock is a key-value pair, where the key is derived from a hash of the Colossus file path, and the value is a unique signature for each operation. This signature consists of a unique identifier for the writer replica and a random nonce to distinguish the operation on that replica. A lock for a file is obtained at the beginning of an operation and released when the operation completes. After acquisition, the writer continuously checks the lock for poisoning. When another operation for the same file is scheduled, it \emph{poisons} the existing lock. Upon detecting the poison, the existing operation is then forced to terminate, allowing the new operation to acquire the lock. Additionally, if a writer terminates without releasing the lock, the new operation seizes the lock after a small delay.

\subsubsection*{Tuning the Caches} We tuned both data and metadata caches for low-latency. These caches operate within the standard shared Borg~\cite{verma2015large} environment, where various jobs compete for resources. Network bandwidth is the most critical resource for the caches due to the high fan out within each cluster. Both data and metadata caches scale horizontally~\cite{rzadca2020autopilot} based on the bandwidth and QPS usage.

\emph{I. Data Cache}: A 1-minute TTL is assigned to the chunks, accommodating most, if not all, consumers. GC is configured to start once the cache reaches 80\% of its capacity. The initial RAM for each replica and the minimum number of replica per cluster is configured to accommodate a minute worth of most recent chunks. The slabs and buckets for key-value pairs are pre-allocated to avoid delays caused by resizing. Chaining is enabled to accommodate overflowing buckets to prevent undesired forced evictions. Furthermore, the data cache utilizes a least-recently modified policy, prioritizing the removal of the oldest written chunk.

\emph{II. Metadata Cache}: The TTL is set to 24 hours which is much larger than the usual rollover time for update files. Since the number of key-value pairs in metadata cache is inherently restricted by the number of files, most capacity and GC settings were deemed unnecessary.

CPU usage on the caches is typically minimal, as most traffic is read-heavy that utilizes a highly efficient software RMA implementation running on Pony Express~\cite{48630}.

To ensure low latency, clients set strict sub-second deadlines on cache operations. Additionally, we tuned the CliqueMap client-side executor for high parallelism by increasing both the number of threads and the buffer space allocated per thread. The underlying execution of operations can be parallelized because we manage concurrency at the application layer. However, as a side-effect, this also increased RAM usage on the consumer side. We then performed manual tuning to optimize for a low resource footprint without impacting latency. In the future, we wish to automate this based on the read traffic pattern. 

Finally, to avoid potential server-side resource conflicts with write operations, reads are deliberately configured to never resort to using RPC mode, except for overflowing bucket reads. However, this still does not prevent OS network buffers and bandwidth conflicts between reads and writes.

\subsubsection*{Fault Tolerance}

The caches can handle single point failures, though multiple failures may impact performance. Both readers and writers can tolerate multiple failures, but this can create temporary imbalance due to operation rescheduling, potentially slowing down operations and increasing latency. To mitigate this, new operations are throttled and existing operation are periodically shed to restore balance.

The scheduler manages the entire life cycle of an operation. During transient periods of scheduler unavailability, such as restart, an operation can continue running in an \emph{orphaned} state thereby maintaining stream continuity. Once the scheduler is back online, it can schedule a new operation for the same file and storage, which will force the existing orphaned operation to terminate, as detailed previously. This straightforward design, inspired by the \emph{make-before-break} strategy, ensures that the stream flow remains uninterrupted.

When scheduler and workers are both unavailable, delays are inevitable. A stream might be disrupted and will need to be re-established.

\section{Evaluation}
\label{sec:evaluation}

In our evaluation, we aim to address several key questions:

\emph{I. Ideal Performance}: How does the system perform under ideal conditions?

\emph{II. Abrupt Load Management}: What happens to the system when an abrupt consumer load is exerted on it?

\emph{III. Fault Tolerance}: How does the system handle faults and ensure resilience?

\emph{IV. Scalability}: How effectively does the storage layer scale in response to egress bandwidth demands?

\emph{V. Backlog Recovery}: How does the system recover from large message backlogs?

All evaluations were performed within the shared Borg~\cite{verma2015large} environment. Each cache replica was allocated 1 CPU core and 8 GB RAM. Consumers were assigned 1 CPU core and 4 GB RAM. Readers and writers were assigned 2 CPU cores and 4 GB RAM, and were configured to scale both vertically and horizontally based on the demand. All jobs were configured to utilize high-priority CPU resources and ran on machines with Pony Express~\cite{48630}, providing highly-efficient software RMA. We used the B4 network~\cite{govindan2016evolve,jain2013b4} for WAN traffic, which is designed for cheaper, low-priority, non-user-facing traffic.

Producers generated messages in data-index format, appending messages into two files: the \emph{data} file, containing message contents, and the \emph{index} file, which primarily stored the offsets of the message contents within the data file. Additionally, producers had a buffering interval of 100ms.

For evaluation purposes, monitoring jobs were turned up alongside consumers. These jobs mimicked consumers and exported performance metrics. The key metric exported by the monitors was \emph{message delivery delay}, calculated as the difference between the producer's time on message production and the consumer's time on message consumption. The delay was comprehensive as it included the time taken by the producer to serialize and write the message, the time taken by the transport layer to copy the message to the destination, and the time taken by the consumer to read and deserialize the message. Although Google's TrueTime~\cite{corbett2013spanner} could have facilitated cross-cluster measurements, we opted to use smeared UTC~\cite{shields2016leap} instead. This decision was based on the observation that clock drifts were negligible compared to the actual delays we encountered. All reported delays have a relative error margin of less than 7\%. Metrics were exported to Monarch~\cite{adams2020monarch} a few minutes into steady-state after startup to avoid capturing noise generated during the initial read up to head.

\pgfplotsset{width=7.44cm,compat=1.16}

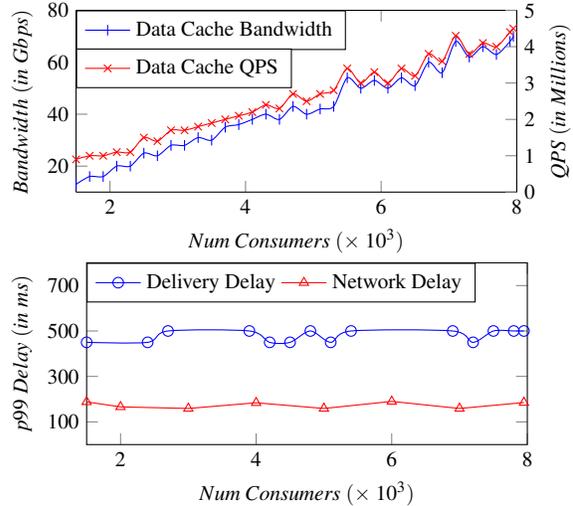
\begin{figure}
\fontsize{8}{10}\selectfont
\begin{tikzpicture}
\begin{axis}[
    xlabel=$Num\ Consumers\ (\times\ 10^3)$,
    ylabel=$Bandwidth\ (in\ Gbps)$,
    xmin=1.5, xmax=8,
    ymin=10, ymax=80,
    axis y line* = left,
    xtick={0,2,4,6,8},
    ytick={20,40,...,80},
    height=4cm]
\addplot[smooth,blue,mark=|] plot coordinates {
    (1.5,13)
    (1.7,16)
    (1.9,16)
    (2.1,20)
    (2.3,20)
    (2.5,25)
    (2.7,24)
    (2.9,28)
    (3.1,28)
    (3.3,31)
    (3.5,30)
    (3.7,35)
    (3.9,36)
    (4.1,38)
    (4.3,40)
    (4.5,38)
    (4.7,43)
    (4.9,40)
    (5.1,42)
    (5.3,43)
    (5.5,54)
    (5.7,50)
    (5.9,53)
    (6.1,50)
    (6.3,54)
    (6.5,51)
    (6.7,60)
    (6.9,56)
    (7.1,68)
    (7.3,62)
    (7.5,66)
    (7.7,63)
    (7.9,68)
    (7.95,70)
};
\pgfplotsset{legend style={at={(0,1)},anchor=north west}}
\addlegendentry{Data Cache Bandwidth}
\end{axis}

\begin{axis}[
    xmin=1.5, xmax=8,
    ymin=0, ymax=5,
    hide x axis,
    axis y line*=right,
    ylabel=$QPS\ (in\ Millions)$,
    ytick={0,1,...,5},
    height=4cm
]

\addplot[smooth,red,mark=x] plot coordinates {
    (1.5,0.9)
    (1.7,1)
    (1.9,1)
    (2.1,1.1)
    (2.3,1.1)
    (2.5,1.5)
    (2.7,1.4)
    (2.9,1.7)
    (3.1,1.7)
    (3.3,1.8)
    (3.5,1.9)
    (3.7,2)
    (3.9,2.1)
    (4.1,2.2)
    (4.3,2.4)
    (4.5,2.3)
    (4.7,2.7)
    (4.9,2.5)
    (5.1,2.7)
    (5.3,2.8)
    (5.5,3.4)
    (5.7,3)
    (5.9,3.3)
    (6.1,3)
    (6.3,3.4)
    (6.5,3.2)
    (6.7,3.8)
    (6.9,3.6)
    (7.1,4.3)
    (7.3,3.8)
    (7.5,4.1)
    (7.7,4)
    (7.9,4.4)
    (7.95,4.5)
};

\pgfplotsset{legend style={at={(0,0.79)},anchor=north west}}
\addlegendentry{Data Cache QPS}
\end{axis}
\end{tikzpicture}

\begin{tikzpicture}[trim axis right]
\begin{axis}[
    xlabel=$Num\ Consumers\ (\times\ 10^3)$,
    ylabel=$p99\ Delay\ (in\ ms)$,
    xmin=1.5, xmax=8,
    ymin=0, ymax=800,
    xtick={2,4,6,8},
    ytick={100,300,500,700},
    height=4cm]
\addplot[smooth,blue,mark=o] plot coordinates {
    (1.5,450)
    (2.4,450)
    (2.7,500)
    (3.9,500)
    (4.2,450)
    (4.5,450)
    (4.8,500)
    (5.1,450)
    (5.4,500)
    (6.9,500)
    (7.2,450)
    (7.5,500)
    (7.8,500)
    (7.95,500)
};

\addplot[solid,red,mark=triangle] plot coordinates {
    (1.5,188)
    (2,166)
    (3,159)
    (4,184)
    (5,159)
    (6,190)
    (7,159)
    (7.95,185)
};

\pgfplotsset{legend columns=-1,legend style={at={(0,1)},anchor=north west}}
\addlegendentry{Delivery Delay}
\addlegendentry{Network Delay}
\end{axis}
\end{tikzpicture}
\caption{Results for experiment 1(a). The fluctuations in data cache bandwidth and QPS are attributed to repeated reads for last partially-filled chunk and the execution of consistent reads when relaxed reads fail. The message delivery delay exported by monitors remained relatively stable regardless of the number of consumers.}
\vspace{-1.5em}
\label{fig:exp1a}
\end{figure}

\subsection{Experiment 1}

In this experiment, we fixed the number of cache replicas. We turned up 9 replicas for the data cache and 6 replicas for the metadata cache in each cluster. To generate workload, we enabled two message streams originating from clusters in the us-central and euro-west regions. These streams were then replicated to 15 destination clusters. The combined producer write rate was 240 Mbps on average. Using r=3.2 and chunk size of 4 KB, this configuration yielded a data cache write QPS of approximately 22,500 in each cluster. The message streams were 120-way sharded, and each shard contained their own data-index file pairs. Each consumer was assigned to read from 4 shards, resulting in a read rate of 8 Mbps per consumer. The polling interval was set to 50ms for readers and 100ms for consumers.

\subsubsection*{Experiment 1(a) - Ideal Case} We started with 1,500 consumers in one of the leaf clusters of the copy tree, which was 3 hops away from each source, located in asia-east region. We scaled-up the number of consumers to see how it affects the latency. The consumers were scaled up smoothly adding 5 replicas every 5 seconds. Throughout the experiment, all upstream clusters reported stable delays. This represents a fairly ideal scenario.

\emph{Observations}: The experiments concluded with 7,950 active consumers. Figure~\ref{fig:exp1a} shows the results. Notably, no fallback to Colossus was observed during stable state of consumers. Peak read bandwidth from the data cache reached 70 Gbps, exceeding the anticipated 62 Gbps by approximately 13\%. A similar trend was noted for QPS, although the deviation from expectations was more pronounced. The peak QPS to the data cache was 4.5M, surpassing the expected 2M by 2.25x. The elevated bandwidth can be attributed to two factors. First, the last partially-filled chunk is read multiple times until it reaches its full size. Second, the chunks that fail relaxed reads are followed up with consistent reads. The latter results in elevated QPS as well.

Monitors consistently reported a p99 message delivery delay of approximately 500ms. This delay stemmed from several factors: producer-side buffering and serial flushing of data and index (120ms), network transit (180ms), and consumer-side periodic polling followed by serial reading of index and data (100ms). Our system added only around 25ms of processing delay along each hop, which included source read I/O, network transmission/receive queuing, and destination write I/O, with the majority of the delay attributed to I/O.

In the following experiments, we capped the number of consumers at 4,000 to ensure we stayed within the resource constraints allocated for each cache replica.

\begin{figure}
\fontsize{8}{10}\selectfont
\begin{tikzpicture}
\begin{axis}[
    xlabel=$Time\ (in\ mins)$,
    ylabel=$Delay\ (in\ ms)$,
    xmin=0, xmax=9,
    ymin=0, ymax=1600,
    axis y line* = left,
    xtick={0,1,...,9},
    ytick={0,400,...,1600},
    height=4cm
]
\addplot[smooth,blue,mark=o] plot coordinates {
    (0,500)
    (1,500)
    (2,500)
    (3,500)
    (4,500)
    (5,500)
    (5.5,500)
    (6.5,1200)
    (7.5,1450)
    (8,500)
    (9,500)
};
\addplot[smooth,red,mark=triangle] plot coordinates {
    (0,5)
    (1,5)
    (2,5)
    (3,5)
    (4,5)
    (5,5)
    (5.5,5)
    (6.5,900)
    (7.5,1100)
    (8,5)
    (9,5)
};
\pgfplotsset{legend style={at={(0,1)},anchor=north west}}
\addlegendentry{Delivery Delay (p99)}
\addlegendentry{Write Latency (p999)}
\end{axis}

\begin{axis}[
    xmin=0, xmax=9,
    ymin=0, ymax=16,
    hide x axis,
    axis y line*=right,
    ylabel=$Fallback\ Rate\ (\times\ 10^3)$,
    ytick={0,4,...,16},
    height=4cm
]

\addplot[dotted,thick,black] plot coordinates {
    (0,0)
    (1,0)
    (2,0)
    (3,0)
    (4,0)
    (5,0)
    (5.5,0)
    (6.5,13.3)
    (7.5,13.3)
    (8,0)
    (9,0)
};

\pgfplotsset{legend style={at={(0,0.63)},anchor=north west}}
\addlegendentry{Fallback}
\end{axis}
\end{tikzpicture}
\caption{Results for experiment 1(b). The elevated cache write latency led to a substantial increase in fallbacks to Colossus by consumers, consequently impacting delivery delay.}
\vspace{-1.5em}
\label{fig:exp1b}
\end{figure}
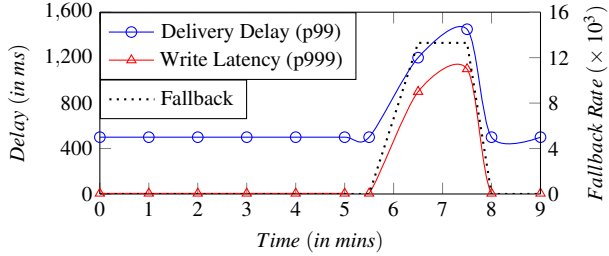

\subsubsection*{Experiment 1(b) - Abrupt Consumer Spike} Utilizing the same cluster setup as in Experiment 1(a), we abruptly scaled the number of consumers from 0 to 4,000 to observe the system's response. This simulates an atypical scenario, like during incident mitigation, where the consumer size surges unexpectedly, putting the system under stress.

\emph{Observations}: The data cache experienced a sharp increase in read QPS, reaching as high as 4.7M, due to numerous requests for trailing bytes of files stored in non-expired chunks. This surge in read QPS created back pressure on writes because both read and write operations shared the same OS network buffers and bandwidth (lacking isolation at that level), temporarily increasing write latency. Consequently, cache operations encountered failures, forcing consumers to fallback to Colossus to retrieve the necessary bytes. This resulted in a noticeable delivery delay increase of approximately 1s for about 150 seconds as shown in Figure~\ref{fig:exp1b}. The system eventually recovered and resumed normal performance.

\begin{figure}
\fontsize{8}{10}\selectfont
\begin{tikzpicture}
\begin{axis}[
    xlabel=$Time\ (in\ mins)$,
    ylabel=$p99\ Delay\ (in\ ms)$,
    xmin=0, xmax=20,
    ymin=0, ymax=2400,
    axis y line* = left,
    xtick={0,2,...,20},
    ytick={0,400,...,2400},
    height=5cm
]
\addplot[smooth,blue,mark=o] plot coordinates {
    (0,200)
    (20,200)
};
\addplot[smooth,red,mark=triangle] plot coordinates {
    (0,250)
    (8,250)
    (9,1200)
    (10,1200)
    (11,250)
    (12,250)
    (20,250)
};
\addplot[smooth,teal,mark=square] plot coordinates {
    (0,400)
    (8,400)
    (9,1400)
    (10,1400)
    (11,400)
    (14,400)
    (15,1200)
    (16,1200)
    (17,400)
    (20,400)
};
\addplot[smooth,black,mark=diamond] plot coordinates {
    (0,600)
    (1,600)
    (2,800)
    (3,600)
    (8,600)
    (9,1600)
    (10,1600)
    (11,600)
    (14,600)
    (15,1400)
    (16,1400)
    (17,600)
    (20,600)
};
\pgfplotsset{legend style={at={(0,1)},anchor=north west}}
\addlegendentry{Source}
\addlegendentry{1st Hop}
\addlegendentry{2nd Hop}
\addlegendentry{3rd Hop}
\end{axis}

\begin{axis}[
    xmin=0, xmax=20,
    ymin=0, ymax=24,
    hide x axis,
    axis y line*=right,
    ylabel=$Operation\ Throttling\ Rate$,
    ytick={0,4,...,24},
    height=5cm
]

\addplot[dashed,thick,red] plot coordinates {
    (0,0)
    (6.8,0)
    (7,6)
    (8,6)
    (9,9)
    (10,0)
};

\addplot[dotted,thick,teal] plot coordinates {
    (0,0)
    (13.8,0)
    (14,2)
    (15,3.5)
    (16,0)
};
\pgfplotsset{legend columns=-1,legend style={at={(0.335,1)},anchor=north west}}
\addlegendentry{1st Hop}
\addlegendentry{2nd Hop}
\end{axis}
\end{tikzpicture}
\caption{Results for experiment 1(c). The solid lines represent p99 delays and non-solid lines represent throttling caused by imbalance. Delay in a hop impacted the delay in downstream hops.}
\vspace{-1.5em}
\label{fig:exp1c}
\end{figure}
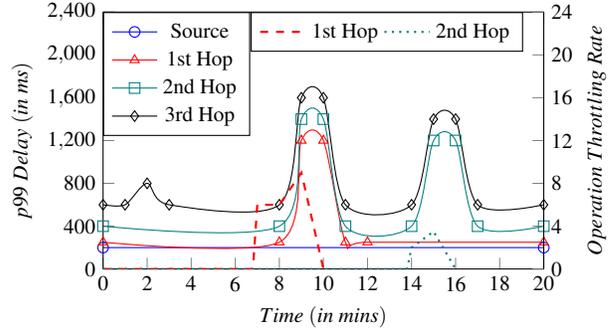

\subsubsection*{Experiment 1(c) - Multi-hop Abrupt Consumer Spike} We repeated experiment 1(b), but along all the hops of the longest branch of the copy tree of one of the message stream. We turned up 4,000 consumers in 4 clusters, including source.

\emph{Observations}: The results are shown in Figure~\ref{fig:exp1c}. We noticed spikes in write latency in the affected clusters, similar to experiment 1(b), which occasionally resulted in operation failures. Although the scheduler rescheduled the operations within 100ms, this rescheduling led to imbalances that caused operation throttling. As a consequence, throttled cache operations triggered Colossus fallbacks. The impact was amplified as a delay in a parent hop affected the entire subtree. The p99 message delivery delay stayed mostly below 600ms, with spikes up to 1.8 seconds coinciding with Colossus fallbacks.

\subsubsection*{Experiment 1(d) - Fault Tolerance} Similar to experiment 1(b), we turned up 4,000 consumers within a single leaf cluster. Additionally, we turned up two backup replicas for both the data cache and the metadata cache. Subsequently, we terminated the cache replicas, followed by the readers and writers, to analyze the resulting impact.

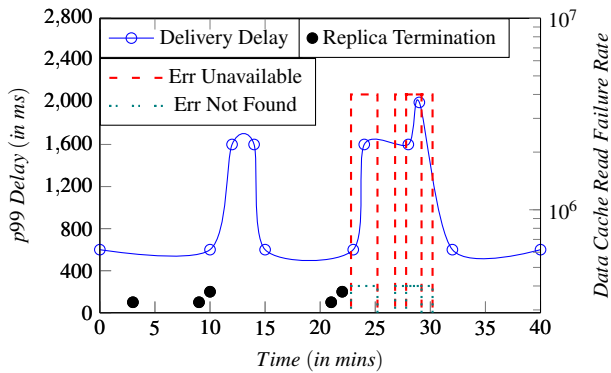
\begin{figure}
\fontsize{8}{10}\selectfont
\begin{tikzpicture}
\begin{axis}[
    xlabel=$Time\ (in\ mins)$,
    ylabel=$p99\ Delay\ (in\ ms)$,
    xmin=0, xmax=40,
    ymin=0, ymax=2800,
    axis y line* = left,
    xtick={0,5,...,40},
    ytick={0,400,...,2800},
    scatter/classes={a={draw=black}},
    height=5.5cm
]

\addplot[smooth,blue,mark=o] plot coordinates {
    (0,600)
    (10,600)
    (12,1600)
    (14,1600)
    (15,600)
    (23,600)
    (24,1600)
    (28,1600)
    (29,2000)
    (32,600)
    (40,600)
};

\addplot[scatter,only marks,scatter src=explicit symbolic]
    table[meta=label] {
    x y label
    3 100 a
    9 100 a
    10 200 a
    21 100 a
    22 200 a
    };

\pgfplotsset{legend style={at={(0,1)},anchor=north west}}
\addlegendentry{Delivery Delay}
\end{axis}

\begin{axis}[
    xmin=0, xmax=40,
    ymin=0, ymax=2800,
    xtick={0,5,...,40},
    ytick={0,400,...,2800},
    axis y line* = left,
    scatter/classes={a={draw=black}},
    height=5.5cm
]

\addplot[scatter,only marks,scatter src=explicit symbolic]
    table[meta=label] {
    x y label
    3 100 a
    9 100 a
    10 200 a
    21 100 a
    22 200 a
    };

\pgfplotsset{legend style={at={(0.93,1)},anchor=north east}}
\addlegendentry{Replica Termination}
\end{axis}

\begin{axis}[
    xmin=0, xmax=40,
    ymin=0, ymax=10000000,
    hide x axis,
    axis y line*=right,
    ylabel=$Data\ Cache\ Read\ Failure\ Rate$,
    ymode=log,
    height=5.5cm
]

\addplot[ybar,red,dashed,thick] plot coordinates {
    (0,0)
    (10,0)
    (12,0)
    (14,0)
    (15,0)
    (15,0)
    (16,0)
    (17,0)
    (18,0)
    (19,0)
    (20,0)
    (24,4000000)
    (28,4000000)
    (29,4000000)
    (32,0)
    (40,0)
};

\addplot[ybar,teal,dotted,thick] plot coordinates {
    (0,0)
    (23,0)
    (24,400000)
    (28,400000)
    (29,400000)
    (32,0)
    (40,0)
};

\pgfplotsset{legend style={at={(0,0.865)},anchor=north west}}
\addlegendentry{Err Unavailable}
\addlegendentry{Err Not Found}

\end{axis}
\end{tikzpicture}
\caption{Results for experiment 1(d). Scatter points represent the specific times when a data cache replica for a particular key-shard was terminated. A significant increase in failures was observed, primarily attributed to key-shard unavailability during the repair process. }
\vspace{-1.5em}
\label{fig:exp1d}
\end{figure}

\emph{Observations}: Data caches demonstrated resilience to single-point failures, maintaining quorum and experiencing no read/write failures when one replica was terminated. However, unavailability and cache operation failures occurred when two replicas of a key-shard were terminated. The backup replica's CPU utilization spiked as it replicated key-value pairs from the remaining replica. This further elongated the unavailability, causing significant Colossus fallbacks of 60,000 reads/seconds and a spike in latency up to 2 seconds as shown in Figure~\ref{fig:exp1d}. A similar pattern emerged for metadata cache when two replicas of a key-shard were terminated, although recovery was faster due to the significantly smaller number of key-value pairs.

In contrast, terminating readers and writers had a negligible impact on latency (not shown), as operations were swiftly rescheduled by the scheduler within 50ms. However, there was a minor effect on load distribution among the remaining replicas, which was gradually resolved through load shedding and operation rescheduling.

\subsection{Experiment 2}

In this experiment, we enabled horizontal scaling for caches, triggered by bandwidth and QPS pressure. We turned up 18 replicas for the data cache and 6 replicas for the metadata cache in one of the leaf cluster. Scaling was configured to trigger whenever a replica breached either 4 Gbps of bandwidth or 400,000 QPS. We simulated a heavier workload by enabling two message streams from clusters in the us-central and euro-west regions. Together, these streams had an average producer write rate of 480 Mbps. The data cache write QPS in each cluster reached approximately 45,000. The message streams were 120-way sharded, and each consumer was assigned 20 shards to read from. Thus, each consumer had a read rate of 80 Mbps. The polling interval was set to 50ms for readers and 500ms for consumers, as a shorter duration led to metadata cache overload.

\begin{figure}[htb]
\fontsize{8}{10}\selectfont
\begin{tikzpicture}
\begin{axis}[
    xlabel=$Num\ Consumers\ (\times\ 10^3)$,
    ylabel=$Bandwidth\ (in\ Tbps)$,
    xmin=0, xmax=20,
    ymin=0, ymax=2.2,
    axis y line* = left,
    xtick={0,2,...,20},
    ytick={0,0.4,...,2},
    height=4cm
]
\addplot[smooth,blue,mark=|] plot coordinates {
    (1,0.09)
    (2,0.180)
    (3,0.270)
    (3.5,0.68)
    (4,0.36)
    (5,0.45)
    (6,0.54)
    (7,0.63)
    (8,0.63)
    (9,0.81)
    (10,0.9)
    (11,0.99)
    (12,1.08)
    (13,1.17)
    (13.5,0.76)
    (14,1.26)
    (15,1.35)
    (16,1.44)
    (16.2,1.3)
    (16.5,1.4)
    (16.7,1.3)
    (16.9,1.4)
    (17,1.53)
    (18,1.62)
    (19,1.71)
    (19.1,1.4)
    (19.3,1.6)
    (19.5,1.3)
    (19.7,1.71)
    (20,1.8)
};
\pgfplotsset{legend style={at={(0,1)},anchor=north west}}
\addlegendentry{Data Cache Bandwidth}
\end{axis}

\begin{axis}[
    xmin=0, xmax=20,
    ymin=0, ymax=110,
    hide x axis,
    axis y line*=right,
    ylabel=$QPS\ (in\ Millions)$,
    ytick={0,20,...,120},
    height=4cm
]

\addplot[smooth,red,mark=x] plot coordinates {
    (1,5)
    (2,10)
    (3,15)
    (3.5,32.5)
    (4,20)
    (5,25)
    (6,30)
    (7,35)
    (8,35)
    (9,45)
    (10,50)
    (11,55)
    (12,60)
    (13,65)
    (13.5,45)
    (14,70)
    (15,75)
    (16,80)
    (16.2,75)
    (16.5,80)
    (16.7,73)
    (16.9,80)
    (17,85)
    (18,90)
    (19,95)
    (19.1,85)
    (19.3,100)
    (19.5,82)
    (19.7,95)
    (20,105)
};

\pgfplotsset{legend style={at={(0,0.79)},anchor=north west}}
\addlegendentry{Data Cache QPS}
\end{axis}
\end{tikzpicture}

\begin{tikzpicture}
\begin{axis}[
    xlabel=$Num\ Consumers\ (\times\ 10^3)$,
    ylabel=$Delay\ (in\ ms)$,
    xmin=0, xmax=20,
    ymin=0, ymax=3000,
    axis y line* = left,
    xtick={0,2,...,20},
    ytick={0,500,...,3000},
    height=5.5cm
]
            
\addplot[smooth,blue,mark=o] plot coordinates {
    (1,1200)
    (2,1200)
    (3,1800)
    (4,1200)
    (5,1200)
    (6,1200)
    (7,1200)
    (8,2200)
    (9,1200)
    (10,1200)
    (11,1200)
    (12,2200)
    (13,2200)
    (14,2200)
    (15,1200)
    (16,2000)
    (17,1200)
    (18,2200)
    (19,2200)
    (20,1200)
};
\addplot[smooth,red,mark=triangle] plot coordinates {
    (1,5)
    (2,5)
    (3,5)
    (3.2,1000)
    (3.4,5)
    (4,5)
    (5,5)
    (6,5)
    (7,5)
    (8,5)
    (9,5)
    (10,5)
    (11,5)
    (12,5)
    (13,5)
    (13.2,1000)
    (13.4,5)
    (14,5)
    (15,5)
    (16,5)
    (16.2,10)
    (16.5,5)
    (16.7,30)
    (16.9,5)
    (17,5)
    (18,5)
    (19,5)
    (19.1,30)
    (19.3,5)
    (19.5,80)
    (19.7,5)
    (20,5)
};
\pgfplotsset{legend style={at={(0,1)},anchor=north west}}
\addlegendentry{Delivery Delay (p99)}
\addlegendentry{Write Latency (p999)}
\end{axis}

\begin{axis}[
    xmin=0, xmax=20,
    ymin=0, ymax=7,
    hide x axis,
    axis y line*=right,
    ylabel=$Fallback\ Rate\ (\times\ 10^5)$,
    ytick={0,1,...,7},
    height=5.5cm
]

\addplot[dotted,thick,black] plot coordinates {
    (1,0)
    (2,0)
    (3,0)
    (3.2,1.90)
    (3.3,1.20)
    (3.4,0)
    (3.5,0.20)
    (3.6,0.40)
    (4,0)
    (5,0)
    (6,0)
    (7,0)
    (7.5,0.40)
    (7.6,0.60)
    (7.7,0.60)
    (7.8,0.70)
    (8,0)
    (9,0)
    (10,0)
    (11,0)
    (12,0)
    (13,0)
    (13.1,2.00)
    (13.2,3.80)
    (13.3,1.50)
    (13.4,5.30)
    (13.5,1.70)
    (14,0)
    (15,0)
    (16,0)
    (16.1,2.00)
    (16.2,0)
    (16.3,2.50)
    (16.4,0)
    (16.5,2.40)
    (16.7,0)
    (17,0)
    (18,0)
    (19,0)
    (19.1,1.30)
    (19.2,2.20)
    (19.3,4.00)
    (19.4,2.00)
    (19.5,6.00)
    (19.6,2.00)
    (19.7,0)
    (20,0)
};

\pgfplotsset{legend style={at={(0,0.77)},anchor=north west}}
\addlegendentry{Fallback}
\end{axis}
\end{tikzpicture}
\caption{Results for experiment 2(a). Dips were observed in data cache bandwidth and QPS, attributed to two factors. First, the spikes in cache write latency. Second, the transient unavailability of the data cache during scale-ups. These dips were accompanied by fallbacks to Colossus.}
\vspace{-1.5em}
\label{fig:exp2a}
\end{figure}
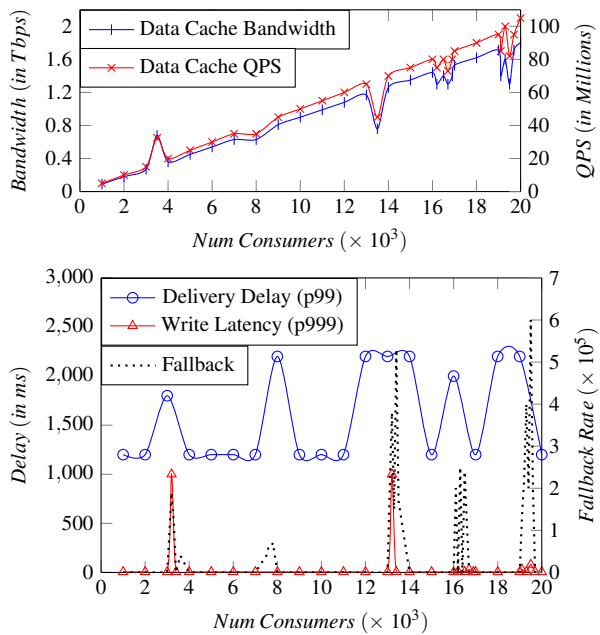

\subsubsection*{Experiment 2(a) - Smooth Scaling} We began with an initial 1,000 consumers and ramped up to 20,000 consumers, incrementally adding 50 replicas every 10 seconds.

\emph{Observations}: The maximum read bandwidth observed from the data cache reached 1.8 Tbps as shown in Figure~\ref{fig:exp2a}. Meanwhile, the metadata cache experienced a peak QPS of 19.2M (20,000 consumers x 2 message streams x 20 shards x 2 files per shard x 2 keys per file x 2 polls per second x 3 replicas). In response to the workload, the data cache scaled up from its initial 18 replicas to 481, surpassing the expected 460 replicas by 4\%. This scale-up led to chunk migrations due to re-sharding, causing transient unavailability of the data cache. This, in turn, forced fallbacks to Colossus, resulting in intermittent drops in data cache bandwidth. Despite this, the p99 message delivery delays exported by monitors remained under 2.5 seconds.

The metadata cache also expanded, growing from 6 replicas to 81, which was 68\% higher than the anticipated 48 replicas. This steeper-than-expected increase in the metadata cache was due to an uneven distribution of key-value pairs, particularly during scale-up events. This imbalance potentially overloaded certain replicas, triggering further scaling. Increasing the replication factor from r=3 to, say, r=10 could address the metadata cache overload issue, but it would come at the cost of increased tail latency during writes. Unfortunately, CliqueMap's current limitations only allowed for a replication factor of 3.

Another interesting observation is the fact that with 27x more data cache replicas after scale-up, we also increased the chunk storage capacity by 27x and so the TTL of chunks could have been as large as 30 minutes. However, in practice, consumers rarely experience such significant delays. We are exploring strategies to vertically scale-down the caches as they are horizontally-scaled.

\subsubsection*{Experiment 2(b) - Abrupt Scaling} We scaled up the consumers suddenly from 0 to 6,500. While such large-scale increases in consumers are infrequent, they can occur in certain clusters during regional outages.

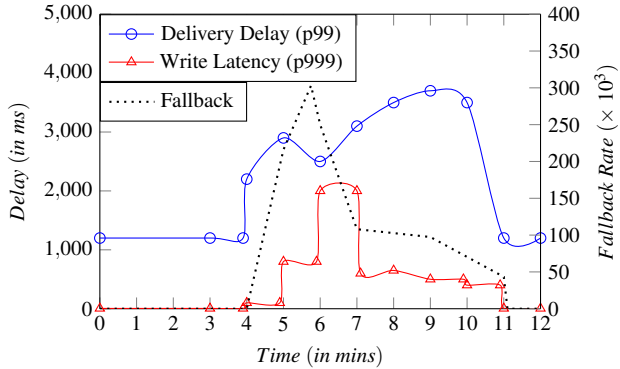
\begin{figure}
\fontsize{8}{10}\selectfont
\begin{tikzpicture}
\begin{axis}[
    xlabel=$Time\ (in\ mins)$,
    ylabel=$Delay\ (in\ ms)$,
    xmin=0, xmax=12,
    ymin=0, ymax=5000,
    axis y line* = left,
    xtick={0,1,...,12},
    ytick={0,1000,...,5000},
    height=5.5cm
]
\addplot[smooth,blue,mark=o] plot coordinates {
    (0,1200)
    (3,1200)
    (3.9,1200)
    (4,2200)
    (5,2900)
    (6,2500)
    (7,3100)
    (8,3500)
    (9,3700)
    (10,3500)
    (11,1200)
    (12,1200)
};
\addplot[smooth,red,mark=triangle] plot coordinates {
    (0,5)
    (3,5)
    (3.9,5)
    (4,100)
    (4.9,100)
    (5,800)
    (5.9,800)
    (6,2000)
    (7,2000)
    (7.1,600)
    (8,650)
    (9,500)
    (9.9,500)
    (10,400)
    (10.9,400)
    (11,5)
    (12,5)
    
};
\pgfplotsset{legend style={at={(0,1)},anchor=north west}}
\addlegendentry{Delivery Delay (p99)}
\addlegendentry{Write Latency (p999)}
\end{axis}

\begin{axis}[
    xmin=0, xmax=12,
    ymin=0, ymax=400,
    hide x axis,
    axis y line*=right,
    ylabel=$Fallback\ Rate\ (\times\ 10^3)$,
    ytick={0,50,...,400},
    height=5.5cm
]

\addplot[dotted,black,thick] plot coordinates {
    (0,0)
    (3,0)
    (4,0)
    (5,216)
    (5.75,303)
    (6,250)
    (7,108)
    (9,97.4)
    (11,43.3)
    (11.1,0)
    (12,0)
};

\pgfplotsset{legend style={at={(0,0.77)},anchor=north west}}
\addlegendentry{Fallback}
\end{axis}
\end{tikzpicture}
\caption{Results for experiment 2(b). The delivery delay was adversely affected for a period of several minutes. A significant number of Colossus fallbacks were observed during this time. Ultimately, the system stabilized and achieved a consistent delivery delay.}
\vspace{-1.5em}
\label{fig:exp2b}
\end{figure}

\emph{Observations}: The peak data cache bandwidth surged to 645 Gbps, and the metadata cache QPS reached 6.24M. This abrupt increase triggered aggressive scaling in both caches, leading to a significant imbalance that necessitated further scaling. The data cache expanded from 18 to 406 replicas, while the metadata cache grew from 6 to 29 replicas. During this period, prolonged cache read and write failures were observed, causing a substantial number of fallbacks to Colossus, peaking at 300,000 reads/second. Consequently, the p99 latency exported by monitors spiked beyond 3 seconds as shown in Figure~\ref{fig:exp2b}. The system eventually recovered after 7 minutes, with delays subsiding to under 1.5 seconds. Once stabilized, the data and metadata caches scaled back down to 171 and 19 replicas respectively, to save resources.

\subsubsection*{Experiment 2(c) - Backlog Recovery} Building on scenario 2(b), we initiated the test with 6,500 consumers, each passively capped at a read rate of 160 Mbps. We then stopped the producer for one of the message stream and started it after a 15-minutes delay. This action resulted in a substantial accumulation of messages awaiting extraction and delivery. Such scenarios can occur when a backlog of messages accumulates due to producer downtime.

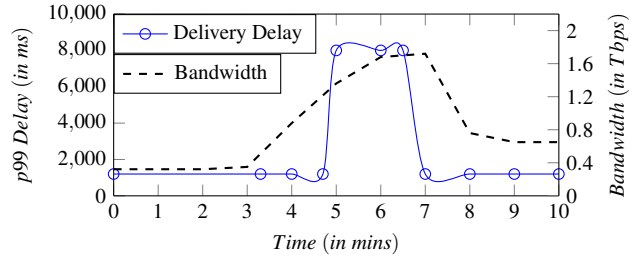
\begin{figure}
\fontsize{8}{10}\selectfont
\begin{tikzpicture}
\begin{axis}[
    xlabel=$Time\ (in\ mins)$,
    ylabel=$p99\ Delay\ (in\ ms)$,
    xmin=0, xmax=10,
    ymin=0, ymax=10000,
    axis y line* = left,
    xtick={0,1,...,10},
    ytick={0,2000,...,10000},
    scaled y ticks=false,
    height=4cm,
    width=7.5cm
]
\addplot[smooth,blue,mark=o] plot coordinates {
    (0,1200)
    (3.3,1200)
    (4,1200)
    (4.7,1200)
    (5,8000)
    (6,8000)
    (6.5,8000)
    (7,1200)
    (8,1200)
    (9,1200)
    (10,1200)
};
\pgfplotsset{legend style={at={(0,1)},anchor=north west}}
\addlegendentry{Delivery Delay}
\end{axis}

\begin{axis}[
    xmin=0, xmax=10,
    ymin=0, ymax=2.2,
    hide x axis,
    axis y line*=right,
    ylabel=$Bandwidth\ (in\ Tbps)$,
    ytick={0,0.4,...,2.2},
    height=4cm,
    width=7.5cm
]

\addplot[dashed,black,thick] plot coordinates {
    (0,0.322)
    (2,0.322)
    (3,0.350)
    (4,0.880)
    (5,1.360)
    (6,1.680)
    (7,1.720)
    (8,0.760)
    (9,0.650)
    (10,0.650)
};

\pgfplotsset{legend style={at={(0,0.78)},anchor=north west}}
\addlegendentry{Bandwidth}
\end{axis}
\end{tikzpicture}
\caption{Results for experiment 2(c). A sudden surge in read bandwidth was observed, which lead to high delivery delay.}
\vspace{-1.5em}
\label{fig:exp2c}
\end{figure}

\emph{Observations}: After the producer restart, the peak bandwidth surged to over 1.6 Tbps as shown in Figure~\ref{fig:exp2c}. This peak was temporary, lasting for only 2 minutes before settling down to 600 Gbps. Due to the transient nature of the spike, the data cache experienced only a minor scale-up from 171 to 184 replicas. The imposed limit on the read rate effectively managed the fan out. However, a significant delivery delay was observed, which eventually resolved once a steady state was reached.

\section{Experiences}
\label{sec:experiences}

\subsubsection*{Development} From early design to deployment, the system required an investment of 8 SWE-years. The development process added 17,500 lines of new, non-test C++ code.

The system underwent extensive testing before production deployment. We run regression tests in the pre-production environment before releasing to production. In production, we monitor the metrics outlined in the evaluation and offer SLOs on latency and cache byte availability.

We have correctness monitors that ensure the bytes read by consumers exactly match the bytes written by producers. While incorrect bytes in compressed message streams would lead to decompression failures, some message streams are uncompressed where it would be particularly concerning if consumers successfully parse malformed messages.

\subsubsection*{Rollout Strategy} Our system spans dozens of clusters, consisting of numerous jobs (with 3 to 400 replicas each) across scheduler, worker, and storage layers. This complexity necessitates a meticulous rollout strategy. During cache rollouts, we adopt a cautious approach, updating only one replica of a given key-shard at a time. After an update, the replicas are given time to repair their state by retrieving key-value pairs from other replicas. 

Worker rollouts cause delay spikes due to throttling and reschedules. Fast rollouts (updating half the replicas at a time) can reduce reschedules but may cause hot-spotting. Conversely, slow rollouts (updating one replica at a time) increases reschedules. We have found that updating one-third of the replicas at once strikes a good balance between hot-spotting and reschedules.

Finally, we never rollout the scheduler and worker jobs together, ensuring scheduler availability during worker rollouts. The rollout strategy discussed so far is slow, sometimes taking several hours. We are continually refining our rollout strategy to strike a better balance between deployment speed and system stability.

\subsubsection*{Integration with Google Ads} The Ads infrastructure relies on an Extract, Transform, and Load~\cite{vassiliadis2009extraction} pipeline. Extraction and transformation are intra-cluster operations primarily consisting of MapReduce~\cite{dean2008mapreduce,zaharia2010spark} jobs. They extract data from various sources such as transactional databases (e.g., Ads F1~\cite{shute2013f1,44686}), data warehouses (e.g., Napa~\cite{50617} and Mesa~\cite{42851}), and real-time log processors (e.g., Ubiq~\cite{45805} and Photon~\cite{41318}), and produce data that is loaded into the serving systems. The extraction and transformation jobs are replicated across a few selected clusters. For instance, each replica of real-time extractors resides in a cluster on a different continent.

\begin{figure}[htb]
\fontsize{8}{10}\selectfont

\begin{tikzpicture}
\begin{axis}[
    xlabel=$Time\ (in\ mins)$,
    ylabel=$Delay\ (in\ ms)$,
    xmin=0, xmax=20,
    ymin=0, ymax=8000,
    axis y line* = left,
    xtick={0,2,...,20},
    ytick={0,2000,...,8000},
    height=4cm,
    width=7.71cm
]
            
\addplot[smooth,blue,mark=o] plot coordinates {
    (0,0)
    (2,4673)
    (4,4941)
    (6,4941)
    (8,3975)
    (12,3975)
    (14,4941)
    (16,3975)
    (18,4941)
    (20,4941)
};
\pgfplotsset{legend style={at={(0,1)},anchor=north west}}
\addlegendentry{Delivery Delay (p99)}
\end{axis}

\begin{axis}[
    xmin=0, xmax=20,
    ymin=0, ymax=7,
    hide x axis,
    axis y line*=right,
    ylabel=$Chunkserver\ Errors\ \%$,
    ytick={0,2,...,8},
    height=4cm,
    width=7.71cm
]
\addplot[dashed,thick,red] plot coordinates {
    (0,0)
    (2,2.63)
    (4,1.683)
    (6,3.8867)
    (8,4.3503)
    (12,3.779)
    (14,3.32)
    (16,4.02)
    (18,3.96)
    (20,3.59)
};

\pgfplotsset{legend style={at={(0.55,1)},anchor=north west}}
\addlegendentry{Err Unavailable}
\end{axis}
\end{tikzpicture}
\caption{Case of data cache unavailability. Colossus chunkservers throttled read requests resulting in unavailability.}
\vspace{-1.5em}
\label{fig:ablation}
\end{figure}
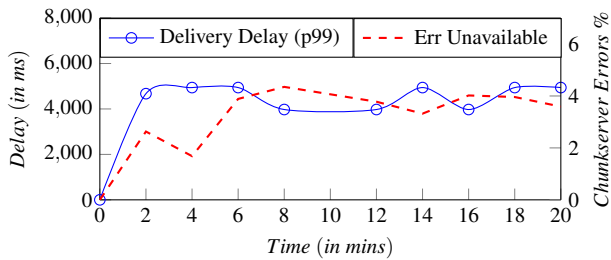

The serving systems, on the other hand, are geographically distributed across dozens of clusters. Each job in the Ads serving system has an in-memory state represented as key-value pairs that is used to serve requests efficiently.

An additional 7,600 lines of non-test C++ code was added for integration with the Ads infrastructure. Each shard of a Google Ads message stream contains two types of entities: \emph{snapshots}, which are compact summaries of a message stream shard up to a logical time~\cite{lamport2019time}, and \emph{updates}, which are continuous, ordered sequences of messages on top of the snapshots. The snapshots are used to build an initial server state and the updates are applied on top of it. Each message updates the value for a particular key. Snapshots are atomically copied across clusters by a batch system, while updates are handled by our system. Update files older than the recent snapshots are garbage-collected.

We migrated several Ads message streams to the new stack. The message delivery delay exported by monitors, measured over a day, shows p95, p99, p999, and p9999 latency of 500ms, 630ms, 730ms, and 5.71s respectively. These latency vary depending on the byte rate of the message stream. The largest message stream experienced a p9999 latency of 8.59s over a day. Even within a single message stream, some shards occasionally experience higher delays. The opportunistic reads in Colossus read operations have 96\% hit rate, saving disk time on the read path.

We have also noticed latency spikes in certain message streams, some lasting several hours. A few of them were caused by bugs, often introduced due to code complexity. Deadlocks have been particularly challenging to debug because of their latent nature and inability of our cluster-local monitors to distinguish between producer-stopped streams and streams stopped due to worker deadlocks. Occasionally, our jobs were scheduled on bad machines, significantly impacting tail latency~\cite{40801} without clear symptoms. In such situations, we resort to rescheduling such jobs based on system metrics. Other delays stemmed from infrequent inter-continental network issues caused by link outages. Although the routing process has automations to address this, the passive nature of the actions still results in delays.

\subsubsection*{Ablation Study} In one scenario, the data cache was unavailable, whereas the metadata cache was available. We recorded a substantial QPS to Colossus chunkservers for several hours. Chunkservers throttle based on CPU, memory, and network time usage, in addition to disk time usage. Consumers experienced throttling resulting in unavailability for 3-5\% of read requests, thereby resulting in a higher delivery delay (Figure~\ref{fig:ablation}). The slight delay intentionally introduced before reads from Colossus (Algorithm~\ref{dfr}) proved effective in managing the QPS thereby resulting in a negligible CPU-related unavailability. The vast majority of unavailability (over 99\%) stemmed from network congestion.

On another occasion, a faulty configuration push rendered both the data and metadata caches unavailable. We observed latency in the order of minutes, as the Colossus file length was polled directly from Bigtable at a much lower frequency.

\subsubsection*{Comparison with Predecessor}  The system before Fast ACS lacked the capability to handle large consumer fanout. With limited fanout, Fast ACS demonstrates greater tail latency stability due to the use of RMA. Furthermore, the adoption of RMA has led to a significant reduction in server-side CPU usage, resulting in cost savings exceeding one-third of the previous total cost.

\section{Key Takeaways}
\label{sec:lessons}

This work yielded several valuable insights for designing scalable, low-latency messaging systems:

1. NIC capacity will be the bottleneck of future network-intensive computing: Given the limited near-term prospects for widespread terabit NIC deployment, achieving ordered data delivery at scale requires designs that efficiently utilize existing NIC capabilities. Our system addresses this by horizontally scaling resources to maximize cluster's NIC capacity utilization.

2. Ordered byte delivery does not need in-order transmission across the network: Instead, parallel out-of-order fetching of data chunks, followed by client-side reassembly, can effectively overcome server-side throughput constraints for low-latency delivery. This concept is already employed in web clients that perform parallel download of file chunks from multiple servers. Our system leverages this principle by chunking message files, pulling these chunks out-of-order, and then assembling and delivering them to consumers in the correct sequence. Through consistent chunk distribution, we also avoid hot-spotting in all scenarios.

3. RMA offers a powerful paradigm for building highly efficient software by minimizing server-side processing overhead: Our implementation strategically utilizes RMA on the client side, reserving server-side computation for fallback scenarios, thus providing significant resource savings.

\section{Related Works}
\label{sec:related}

Distributed storage systems are evolving to leverage modern hardware capabilities. For example, NVM- and RDMA-aware HDFS~\cite{islam2016high} provides remote SSD reads for file bytes. Octopus~\cite{zhu2021octopus+} enhances this further with low-latency file metadata access. Colossus also supports similar functionality with SSDs and CliqueMap, too, can use SSDs as the underlying storage. However, SSDs are not ideal for our needs for several reasons. First, their lifespan would be limited due to wear-leveling caused by continuous writes. Second, they offer lower read throughput and higher latency compared to RAM. Third, our hot storage requirements are minimal, not justifying the capacity of SSDs. 

Tachyon~\cite{li2014tachyon} is an RMA-based in-memory file system available for big data processing. However, it is not optimized for tail reading due to its large block sizes, which can lead to hot-spots on the tail block servers. Reducing the block size can help, but it may lead to latency hits on read path since block metadata access still relies on RPCs. Our approach of distributing 4 KB chunks as key-value pairs significantly reduces hot-spotting during tail reads.

Memcache~\cite{nishtala2013scaling} is a popular productionized distributed in-memory key-value store. It provides low-latency storage and lookup and integrates well with external sharding engines like Slicer~\cite{adya2016slicer} to offer scalability. However, we decided to use CliqueMap due to two reasons. First, CliqueMap supports replication out-of-the-box, which makes it resilient to a single point of failure and makes our rollout process simple, unlike Memcache, which requires certain wrappers to support replication. Second, CliqueMap makes use of RMA on the read path, achieving significantly higher throughput and saving valuable server-side CPU while maintaining a comparable median latency. Other RMA key-value systems like FaRM-KV~\cite{dragojevic2014farm} could provide serialized consistent lookups using primary/backup architecture; however, serialization-based consistency was not required in our design, and so we opted for a simpler quorum-based system. Pilaf~\cite{mitchell2013using} is one such RMA KV-store with quorum-based support and resembles performance similar to CliqueMap. However, CliqueMap is already productionized at Google and serves as a backend for critical infrastructures in Ads, Maps, and YouTube.

Lastly, while there is extensive research on userspace file systems like FUSE~\cite{vangoor2017fuse} and its extension, extFUSE~\cite{bijlani2019extension}, which provide a VFS-like interface for creating custom file systems, we chose not to use them. First, our needs were limited to dealing with files, not directories. This allowed us to utilize hash-based addressing instead of a hierarchical file system and a key-value store was a natural fit. Second, our caching file system was essentially a reflection of the underlying Colossus file system supporting append-only files. Implementing a full file system was unnecessary.

\section{Future Work}
\label{sec:future}

To ensure fair and efficient caching for a growing number of message streams, we are exploring key-space isolation to prevent resource contention. Currently, the shared data cache can be overused by individual streams, impacting others. By providing dedicated cache space, each stream can independently manage its usage and chunk expiration, improving overall system stability and fairness.

Each Google cluster has a capacity limit ranging from hundreds to tens of thousands of machines. During experiment 2, we observed that multiple consumers were running on the same machines, sometimes up to 10. One effective technique to limit intra-cluster bandwidth would be to implement a sidecar ambassador process~\cite{burns2016design} that retrieves messages onto local machines and relays them to consumers. This approach would allow us to scale up to millions of consumers per cluster while enforcing a limit on anticipated data cache bandwidth and metadata cache QPS.

In experiment 2, we observed that the primary factor driving up the latency was the increased consumer polling interval. Reducing the polling interval can help bring down latency but can cause QPS limits breach on the metadata cache. There can be many strategies to mitigate the QPS problem. First, the bandwidth utilization per metadata cache replica is low and so the network is essentially non-constraining; the QPS limits can be raised by increasing the number of software RMA endpoints. Second, because the write QPS is much lower than read QPS, we can combine metadata for multiple files into a single key-value pair. This combination can be automated based on the consumer read pattern.

Finally, we are investigating the potential of  RMA and other network enhancements to improve performance in broader scenarios, such as cross-cluster RMA-based reads. However, current network deployment limitations prevent us from supporting this functionality in all cases.

\section{Conclusion}
\label{sec:conclusion}

This paper presents the design of Fast ACS, a file-based ordered message delivery system. At its core is a multi-layer storage that has been demonstrated to be a valuable component, helping the system achieve Tbps-scale consumer fan-out per cluster and deliver messages with minimal latency. This system has been widely deployed in production.

\section*{Acknowledgments}

We gratefully acknowledge contributions to the design and implementation of the system from many colleagues at Google. Special mention for Aman Shaikh, Apar Madan, Ayush Jha, Ben T. Harper, Bin Lyu, Chen-Han Ho, Connor Quagliana, Erin Rovelstad, Florentina I. Popovici, Hongchen Li, Lily Chen, Manav Prajapati, Mengjie Xia, Nick Golob, Paul D. Bartlett, Serafi Zanikolas, Sucharitha Vasudevan, Tao Huang, and Youer Pu. We also thank Jeff Mogul, Tao Huang, and Ashish Gupta for their helpful suggestions on the presentation of the work. Finally, we thank the anonymous reviewers and our shepherd, Fernando Pedone, for their feedback.


\bibliographystyle{plain}
\bibliography{\jobname}

\renewcommand{\thesection}{\Alph{section}}
\setcounter{section}{0}

\clearpage
\newpage
\mbox{~}
\section{Formal Specification for Dueling Writers}
\label{sec:appendix}

We developed a formal specification for the system, which consists of multiple writers for the cache file and a single writer for the corresponding Colossus file. The system was modeled using TLA+~\cite{lamport2002specifying}. The goal was to check for correctness of the system, manifested in the following way:

\begin{itemize}
\item
\emph{Safety}: Consumers receive bytes strictly in the order they were produced. This is formally expressed by the following invariant:

\emph{{Invariant} == }

\emph{$\ \ \ \ \ \ \ \ \ \land\ readBytes = [i \in 1..Len(readBytes) \mapsto kBytes[i]]$}

where \emph{readBytes} are the bytes read by the consumer and \emph{kBytes} are the bytes written by the producer in source.

\item
\emph{Eventual Progress}: The consumers eventually receives all the bytes. Formally, the following property is satisfied:

\emph{{Termination} ==}

\emph{$\ \ \ \ \ \ \  \diamondsuit \ ((pc = ``done") \implies (readBytes = kBytes))$}

where \emph{pc} = \emph{"done"} is the termination marker.

\end{itemize}

The system's safety is intuitively clear: data cache chunks always contain the correct bytes, if present, corresponding to the file positions, and each chunk write is atomic. The corresponding Colossus file serves as a safety net for partially-filled or missing chunks.

Eventual progress is ensured because the Colossus file continually grows in length, eventually forcing consumers to fall back and fetch bytes directly from Colossus.

We used the TLA+ Model Checker~\cite{yu1999model} to verify the specification under two configurations:

\begin{itemize}
\item \emph{Configuration 1}: 4 chunks, each with 4 bytes, and 2 cache writers. This resulted in 85,690,897 distinct states.

\item \emph{Configuration 2}: 2 chunks, each with 2 bytes, and 4 cache writers. This resulted in 12,062,699 distinct states.
\end{itemize}

In both runs, the model checker successfully verified the system's safety and eventual progress.



\section*{Supplementary material}

―――――――――― MODULE *FastACS* ――――――――――

This module offers a formal specification for Fast *ACS*, focusing on the
verification of system correctness in presence of multiple, dueling writers.
The components it models include:

1. Data cache: This stores file bytes, divided into *kNumChunks chunks*, each
   of size *kChunkSize*
2. *Metadata cache* : *This stores the length of the data cache file.*
3. *Colossus* : *This stores the bytes in a single file.*
4. *Cache writers* : *These handle the writing of bytes to the data cache. A
   single cache file can have multiple writers specified by the parameter
   kNumCacheWriters*
5. *Colossus writer* : *This is responsible for writing bytes to the Colossus
   file.*
6. *Consumer* : *This reads bytes, prioritizing the data cache and using the
   Colossus file as a fallback when necessary.*

*A write to the data cache and the metadata cache form a non − atomic action,
represented as a series of steps. In contrast, a write to Colossus is atomic,
meaning that both bytes and metadata are written atomically, and this is
modeled as a single, atomic step.*

*The following parameters are required to run the model* :

1. *kChunkSize* : *The size of each chunk in the data cache.*
2. *kNumChunks* : *The number of chunks in the data cache.*
3. *kNumCacheWriters* : *The number of cache writers.*

*Verifications performed* :

*Safety* : *The consumer receive bytes strictly in the order they were produced.*
*Liveness* : *The consumer eventually receives all the bytes.*

EXTENDS *Naturals*, *Sequences*, *TLC*
CONSTANTS *kChunkSize*, *kNumChunks*, *kNumCacheWriters*
VARIABLE
   *Program counter. Represented as a string that can be* "start" *or* "done".
   *pc*,
   *Chunks in the data cache. Represented as a map from chunk index to the
   sequence of bytes in that chunk. 0 − indexed.*
   *dataCacheChunkBytes*,
   *Length of the data cache in the metadata cache. Represented as an
   integer.*
   *metadataCacheLength*,

1

*Bytes in the Colossus file. Represented as a sequence of bytes.*
$colossusBytes,$
*Next write position for each cache writer. Represented as an array of integers.*
$cwWritePos,$
*Last partial chunk for each cache writer. Represented as an array of sequence of bytes.*
$cwPartialChunk,$
*Position till which metadata cache write is complete for each cache writer. Represented as an array of integers.*
$cwMetadataCachePos,$
*Best known file position to consumers. Represented as an integer.*
$bestPos,$
*Bytes read by the consumer. Represented as a sequence of bytes.*
$readBytes$

*File size.*
$kSize \triangleq kChunkSize * kNumChunks$
*File bytes. This is a sequence of kSize natural numbers.*
$kBytes \triangleq [i \in 1 \,.\, kSize \mapsto i]$

*Initial state.*
$Init \triangleq$

*Program counter.*

$\land pc = \text{``start''}$

*Storage state.*

*Chunks in the data cache. Each chunk is a sequence of bytes. $0-indexed$.*
$\land dataCacheChunkBytes =$
$\quad [k \in (0 \,.\, kNumChunks - 1) \mapsto \langle\rangle]$
*Length of the data cache in the metadata cache.*
$\land metadataCacheLength = 0$
*Colossus file bytes.*
$\land colossusBytes = \langle\rangle$

*Cache writer state (distinguished by writer id).*

*Next write position.*
$\land cwWritePos = [id \in (1 \,.\, kNumCacheWriters) \mapsto 0]$
*Last partial chunk bytes.*
$\land cwPartialChunk = [id \in (1 \,.\, kNumCacheWriters) \mapsto \langle\rangle]$

*Position till which metadata cache write is complete.*
∧ $cwMetadataCachePos = [id \in (1 \ldots kNumCacheWriters) \mapsto 0]$

*Consumer state.*

*Best known position.*
∧ $bestPos = 0$
*Read bytes.*
∧ $readBytes = \langle \rangle$

*A cache writer with identifier id writes a single byte to the data cache,*
*iff a byte is pending to be written.*
$DataCacheWrite(id) \triangleq$
 ∧ $dataCacheChunkBytes' =$
  IF $cwWritePos[id] = kSize$
   THEN $dataCacheChunkBytes$
  ELSE
   *Appends the next byte to the partial chunk and updates the chunk in*
   *the data cache.*
   $[dataCacheChunkBytes$ EXCEPT
    $![cwWritePos[id] \div kChunkSize]$
     $= Append(cwPartialChunk[id],$
       $kBytes[cwWritePos[id] + 1])]$
 ∧ $cwPartialChunk' =$
  IF $(cwWritePos[id] = kSize) \lor (Len(cwPartialChunk[id]) + 1 = kChunkSize)$
   THEN $[cwPartialChunk$ EXCEPT $![id] = \langle \rangle]$
  ELSE
   *Appends the next byte to the partial chunk.*
   $[cwPartialChunk$ EXCEPT
    $![id] = Append(cwPartialChunk[id], kBytes[cwWritePos[id] + 1])]$
 ∧ $cwWritePos' =$
  IF $cwWritePos[id] = kSize$
   THEN $cwWritePos$
  ELSE
   $[cwWritePos$ EXCEPT $![id] = cwWritePos[id] + 1]$
 ∧ UNCHANGED
  $\langle pc,$
   $metadataCacheLength, colossusBytes,$
   $cwMetadataCachePos,$
   $bestPos, readBytes \rangle$

*A cache writer with identifier id writes the current position to the*
*metadata cache, iff the current position has been updated.*
$MetadataCacheWrite(id) \triangleq$
 ∧ $metadataCacheLength' =$

```
        IF cwMetadataCachePos[id] < cwWritePos[id]
          THEN cwWritePos[id]
        ELSE
          metadataCacheLength
  ∧ cwMetadataCachePos′ =
      IF cwMetadataCachePos[id] < cwWritePos[id]
        THEN [cwMetadataCachePos EXCEPT ![id] = cwWritePos[id]]
      ELSE
        cwMetadataCachePos
  ∧ UNCHANGED
      ⟨pc,
       dataCacheChunkBytes, colossusBytes,
       cwWritePos, cwPartialChunk,
       bestPos, readBytes⟩
```

*The Colossus writer appends a single byte to Colossus file, iff a byte is pending to be written.*

```
ColossusWrite ≜
  ∧ colossusBytes′ =
      IF Len(colossusBytes) = kSize
        THEN colossusBytes
      ELSE
        Append(colossusBytes, kBytes[Len(colossusBytes) + 1])
  ∧ UNCHANGED
      ⟨pc,
       dataCacheChunkBytes, metadataCacheLength,
       cwWritePos, cwPartialChunk, cwMetadataCachePos,
       bestPos, readBytes⟩
```

*The consumer reads the data cache length from the metadata cache and updates the best position.*

```
MetadataCacheRead ≜
  ∧ bestPos′ =
      IF bestPos < metadataCacheLength
        THEN metadataCacheLength
      ELSE
        bestPos
  ∧ UNCHANGED
      ⟨pc,
       dataCacheChunkBytes, metadataCacheLength, colossusBytes,
       cwWritePos, cwPartialChunk, cwMetadataCachePos,
       readBytes⟩
```

*The consumer reads the length of the Colossus file and updates the best position.*

$ColossusLengthRead \;\triangleq$

$\land\ bestPos' =$
  IF $bestPos < Len(colossusBytes)$
    THEN $Len(colossusBytes)$
  ELSE
    $bestPos$
$\land$ UNCHANGED
  $\langle pc,$
  $dataCacheChunkBytes, metadataCacheLength, colossusBytes,$
  $cwWritePos, cwPartialChunk, cwMetadataCachePos,$
  $readBytes\rangle$

*The consumer attempts to reads a single byte from the data cache(preferably) and Colossus file(as fallback).*

$DataRead\ \triangleq$
  $\land\ readBytes' =$
    IF $(Len(readBytes) = kSize) \lor (Len(readBytes) = bestPos)$
      THEN $readBytes$
    ELSE IF $Len(dataCacheChunkBytes[Len(readBytes) \div kChunkSize])$
              $> (Len(readBytes)\%kChunkSize)$
      *Read the next byte from the data cache.*
      THEN $Append(readBytes,$
              $dataCacheChunkBytes[$
                $Len(readBytes) \div kChunkSize][$
                  $(Len(readBytes)\%kChunkSize) + 1])$
    ELSE IF $Len(colossusBytes) > Len(readBytes)$
      *Read the next byte from the Colossus file.*
      THEN $Append(readBytes, colossusBytes[Len(readBytes) + 1])$
    ELSE
      $readBytes$
  $\land$ UNCHANGED
    $\langle pc,$
    $dataCacheChunkBytes, metadataCacheLength, colossusBytes,$
    $cwWritePos, cwPartialChunk, cwMetadataCachePos,$
    $bestPos\rangle$

$Finish\ \triangleq$
  $\land\ pc' =$
    IF $Len(readBytes) = kSize$
      THEN "done"
    ELSE
      $pc$
  $\land$ UNCHANGED
    $\langle dataCacheChunkBytes, metadataCacheLength, colossusBytes,$
    $cwWritePos, cwPartialChunk, cwMetadataCachePos,$
    $bestPos, readBytes\rangle$

$Next \triangleq$
  $\lor \exists w \in 1 \mathbin{\ldotp\ldotp} kNumCacheWriters : DataCacheWrite(w)$
  $\lor \exists w \in 1 \mathbin{\ldotp\ldotp} kNumCacheWriters : MetadataCacheWrite(w)$
  $\lor ColossusWrite$
  $\lor MetadataCacheRead$
  $\lor ColossusLengthRead$
  $\lor DataRead$
  $\lor Finish$

$Spec \triangleq$
  $\land Init$
  $\land \Box[Next]$
    $\langle pc,$
    $dataCacheChunkBytes, metadataCacheLength, colossusBytes,$
    $cwWritePos, cwPartialChunk, cwMetadataCachePos,$
    $bestPos, readBytes \rangle$

*Invariants throughout the execution of the system.*
$Invariants \triangleq$
  $\land metadataCacheLength \in (0 \mathbin{\ldotp\ldotp} kSize)$
    *Colossus bytes are a prefix of producer bytes.*
  $\land colossusBytes = [i \in 1 \mathbin{\ldotp\ldotp} Len(colossusBytes) \mapsto kBytes[i]]$
  $\land bestPos \in (0 \mathbin{\ldotp\ldotp} kSize)$
    *Bytes read by the consumer are a prefix of producer bytes (strict*
    *in − ordering).*
  $\land readBytes = [i \in 1 \mathbin{\ldotp\ldotp} Len(readBytes) \mapsto kBytes[i]]$
    *Bytes are not read past the best known position.*
  $\land Len(readBytes) \leq bestPos$

*Termination condition: When the system is done, all produced bytes would be*
*read by the consumer.*
$Termination \triangleq$
  $\Diamond((pc = \text{``done''}) \Rightarrow (readBytes = kBytes))$

---

THEOREM  $Spec \Rightarrow \Box Invariants$
THEOREM  $Spec \Rightarrow \Box Termination$



\end{document}